%% file: main.tex
\documentclass[twocolumn]{aastex63}

\usepackage{colortbl} 
\usepackage{multirow}
\usepackage{verbatim}
\usepackage{comment} 
\usepackage{amsmath}
\usepackage{array}

\usepackage{rotating}
\usepackage{enumitem}
\usepackage{natbib}
\usepackage{graphics,graphicx}
\usepackage{subcaption}
\usepackage{caption}
\usepackage[capitalise]{cleveref}
\usepackage{hyperref}

\makeatletter
\def\@hex@@Hex#1%
 {\if a#1A\else \if b#1B\else \if c#1C\else \if d#1D\else
  \if e#1E\else \if f#1F\else #1\fi\fi\fi\fi\fi\fi \@hex@Hex}
\makeatother

\definecolor{asparagus}{rgb}{0.53, 0.90, 0.52}
\definecolor{darkseagreen}{rgb}{0.56, 0.74, 0.56}
\definecolor{seagreen}{rgb}{0.18, 0.55, 0.34}
\definecolor{bostonuniversityred}{rgb}{0.8, 0.0, 0.0}
\definecolor{babyblue}{rgb}{0.54, 0.81, 1.}
\definecolor{beaublue}{rgb}{0.74, 0.83, 0.9}
\definecolor{ceruleanblue}{rgb}{0.16, 0.32, 0.75}
\definecolor{smalt(darkpowderblue)}{rgb}{0.0, 0.2, 0.6}

\newcommand{\operon}{\textsc{operon}}
\newcommand{\skirt}{\textsc{skirt}}
\newcommand{\Av}{A_{\rm V}}
\newcommand{\lamv}{\lambda_{\rm V}}
\newcommand{\sfrunit}{{\rm M_{\odot} \, yr^{-1}}}

\newcommand{\splitatcommas}[1]{%
  \begingroup
  \begingroup\lccode`~=`, \lowercase{\endgroup
    \edef~{\mathchar\the\mathcode`, \penalty0 \noexpand\hspace{0pt plus 1em}}%
  }\mathcode`,="8000 #1%
  \endgroup
}

\revised{\today}
\graphicspath{{./}{figures/}}

\begin{document}

\include{definitions}

\title{Learning the Universe: The Structure of Dust Attenuation Curves in Galaxy Simulations}
\shorttitle{Learning the Universe: Dust Attenuation Curves}
\shortauthors{Sommovigo, Bartlett, et al.}

\author[0000-0002-2906-2200]{Laura Sommovigo}
\altaffiliation{These authors contributed equally to this work.}
\affiliation{Department of Astronomy, Columbia University, New York, NY 10027, USA}
\affiliation{Center for Computational Astrophysics, Flatiron Institute, 162 Fifth Avenue, New York, NY 10010, USA}
\begingroup
\renewcommand{\thefootnote}{}
\footnotetext{\href{mailto:laura.sommovigo.work@gmail.com}{laura.sommovigo.work@gmail.com}}
\addtocounter{footnote}{-1}
\endgroup

\author[0000-0001-9426-7723]{Deaglan J. Bartlett}
\altaffiliation{These authors contributed equally to this work.}
\affiliation{Astrophysics, University of Oxford, Denys Wilkinson Building, Keble Road, Oxford OX1 3RH, UK}
\begingroup
\renewcommand{\thefootnote}{}
\footnotetext{\href{mailto:deaglan.bartlett@physics.ox.ac.uk}{deaglan.bartlett@physics.ox.ac.uk}}
\addtocounter{footnote}{-1}
\endgroup

\author[0000-0001-8855-6107]{Rachel K. Cochrane}
\affiliation{Jodrell Bank Centre for Astrophysics, University of Manchester, Oxford Road, Manchester M13 9PL, UK}

\author[0000-0003-3207-8868]{Matthew Ho}
\affiliation{Department of Astronomy, Columbia University, New York, NY 10027, USA}


\author[0000-0001-7964-5933]{Christopher C. Lovell}
\affiliation{Kavli Institute for Cosmology, University of Cambridge, Madingley Road, Cambridge CB3 0HA, UK}
\affiliation{Institute of Astronomy, University of Cambridge, Madingley Road, Cambridge CB3 0HA, UK}

\author[0000-0002-6748-6821]{Rachel S.~Somerville}
\affiliation{Center for Computational Astrophysics, Flatiron Institute, 162 5th Avenue, New York, NY 10010, USA}

\begin{abstract}
\noindent Dust attenuation is a major source of systematic uncertainty in both SED fitting and forward modeling of galaxy populations, yet the functional form used to parameterize attenuation curves has received surprisingly little systematic scrutiny. Particular unanswered questions include: how many free parameters are genuinely needed, and which analytic expression best captures the full diversity of attenuation curve shapes in galaxies across cosmic time?
Using a large library of synthetic attenuation curves from TNG50 and TNG100 galaxies post-processed with the 
\skirt{} radiative transfer code using three dust mixtures (Milky Way, SMC, and stellar dust), 
we show via Information-Ordered Bottleneck analysis that \textit{exactly four parameters} 
are needed to capture the diversity of attenuation curves. 
Guided by this result, we use symbolic regression to derive a new, interpretable four-parameter 
attenuation model that outperforms existing parameterizations in recovering both attenuation curves 
and emergent fluxes across all dust mixtures explored.
The four parameters of this model have clear physical interpretations: UV bump strength, FUV slope, UV-bump transition curvature, and large-scale optical slope. Their correlations with galaxy properties are primarily regulated by star-formation 
rate surface density, metallicity, and stellar--dust geometry, and are largely preserved across dust mixtures -- 
except for the bump-sensitive parameters, which retain a stronger dependence on grain composition. 
We further provide symbolic-regression scaling relations linking all four parameters to quasi-observable galaxy properties, 
offering a physically motivated route to assign realistic attenuation curves in SED fitting and forward modeling 
without radiative-transfer calculations.
\vspace{22pt}
\end{abstract}

\section{Introduction \label{sec:intro}}

The spectral energy distributions (SEDs) of galaxies are fundamentally shaped by dust, which absorbs and scatters stellar UV and optical photons -- reducing the observed flux at short wavelengths -- and re-emits the absorbed energy in the infrared \citep{Draine89,Meurer99,Calzetti00,Draine03}. This paper is concerned with the former effect: the wavelength-dependent absorption and scattering of light, encoded in the attenuation curve. Accurately accounting for this is essential both for the \textit{inference} of galaxy physical properties -- including stellar mass, SFR, and stellar ages -- from photometric or spectroscopic observations via SED fitting \citep{Salim20,DustE22,Markov23,Fisher25,Chworowsky26}, and for the \textit{forward modeling} of galaxy populations in observable space \citep[e.g.][]{Somerville2012,Lacey2016,Lagos18,Trayford15,Narayanan18,Hahn2022,Mauerhofer23,Cochrane24,Lovell24,Sommovigo_2025}. In both cases, the accuracy of the resulting inference or prediction is fundamentally limited by how well the wavelength-dependent impact of dust -- encoded in the attenuation curve -- is understood and parameterized.

In this context, it is important to distinguish between \textit{extinction} and \textit{attenuation} \citep[see][for an extensive review on this topic]{Salim20}. Extinction refers to the absorption and scattering of photons along a single line of sight toward an individual star through a foreground dust screen, and is a local quantity. It depends on the dust column density and the intrinsic properties of the dust grains (e.g.\ chemical composition and size distribution) along that sightline \citep{Cardelli89,Fitzpatrick99,Gordon03}, with scattering into the line of sight typically playing a sub-dominant role. 
Attenuation curves -- the focus of this work -- instead describe the net effect of dust on the integrated emission from patches of and/or entire galaxies \citep{Calzetti94,Calzetti00}, which depends not only on grain properties but also on the relative geometry of stars and dust, the optical depth of the medium, and the mix of stellar populations \citep{Narayanan18,Matsumoto26}. Concretely, the attenuation at wavelength $\lambda$ is defined as:
\begin{equation}
A_\lambda = -2.5\,\log_{10}\!\left(\frac{F_{\lambda,\mathrm{obs}}}{F_{\lambda,\mathrm{int}}}\right),
\end{equation}
where $F_{\lambda,\mathrm{obs}}$ and $F_{\lambda,\mathrm{int}}$ are the observed and intrinsic fluxes, respectively. Attenuation curves are typically normalized either to the $V$-band attenuation, $A_\lambda/\Av$, or to the color excess $E(B-V)$, yielding $k(\lambda) = A_\lambda / E(B-V)$, with $k(\lambda)/R_V = A_\lambda/\Av$ and $R_V = \Av/E(B-V)$.

In our own Galaxy, extinction curves have been measured along individual lines of sight towards stars using the ``pair method'' \citep{Cardelli89,Fitzpatrick99}, revealing a characteristic rise from infrared (IR) to far-ultraviolet (FUV) wavelengths and a prominent absorption feature at $2175\,\angstrom$, attributed to small ($\simlt 10^{-3}\,{\rm \mu m}$) carbonaceous grains \citep{Cardelli89}. Similar measurements of the Magellanic Clouds reveal a diversity of extinction curve shapes: the LMC curve exhibits a weak $2175\,\angstrom$ bump and a steep FUV rise, while the SMC curve is nearly featureless and steeper still \citep{Koornneef81,Nandy81,Prevot84,Bouchet85}. 
In galaxies where individual stars cannot be resolved, such direct measurements are not possible, and one must instead rely on statistical or population-level methods.

The foundational observational work on galaxy attenuation curves is that of \citet{Calzetti00}, who derived an average attenuation law from UV-optical spectra of $39$ local starburst galaxies, assuming a foreground dust screen geometry. The resulting law is notably flatter (``grayer'') than Milky Way or Magellanic Cloud extinction curves, lacks the $2175\,\angstrom$ feature entirely, and has $R_V \simeq 4.05$. It has since been widely adopted in SED fitting and applied to galaxy samples up to $z \sim 1$--$3$ \citep{Reddy15,Salmon16,Battisti20,Shivaei22}.

More recent analyses of large samples have shown that attenuation curves are far from universal. Studies of $\sim 230{,}000$ galaxies at $z \sim 0$ \citep{Salim18} find a broad range of slopes and UV bump amplitudes, with more optically thick (and generally more massive) galaxies displaying flatter curves and bump strengths spanning from zero to MW-like values (sample-averaged $\sim 1/3$ that of the MW). At $z \sim 1.4$--$2.6$, analyses of data from the MOSFIRE Deep Evolution Field (MOSDEF) Survey \citep{Reddy15,Shivaei20,Shivaei22} reveal a positive correlation between metallicity and both slope and bump amplitude, suggesting that more evolved, metal-rich galaxies retain more carbonaceous dust and exhibit shallower curves.

The advent of the James Webb Space Telescope (\textit{JWST}; \citealt{JWST_06,JWST_23}) has extended these studies out to $z \sim 10$. The $2175\,\angstrom$ feature has now been reported in individual galaxies at $z > 6$ \citep{Witstok23,Markov24,Lin25}, indicating early and efficient dust production and ISM reprocessing, while statistical analyses of UV-bright galaxies at $z \sim 4$--$10$ are beginning to constrain population-level evolution of attenuation curve properties \citep{Markov23,Markov24,Fisher25,Ormerod25,Shivaei25,Chworowsky26,Rodighiero26}. Collectively, these studies hint at flatter/grayer curves and weaker bumps at high-$z$, though a coherent picture has yet to emerge, in part because the highest-$z$ samples suffer from low number statistics and a UV-bright selection. The latter might favor galaxies with a significant fraction of unobscured young stars, which — by contributing UV flux without attenuation — can itself flatten observed attenuation curves even at fixed grain composition and in the absence of high optical depth sightlines \citep{Narayanan18, Matsumoto26, Cochrane24, Sommovigo26}.

Interpreting the observed diversity of attenuation curves and their correlations with galaxy properties requires understanding physical processes operating across a vast range of scales, which no current galaxy simulation can fully resolve without sub-grid modeling. At the microscopic level, the underlying extinction curve is set by the dust grain size distribution and chemical composition: small carbonaceous grains, including polycyclic aromatic hydrocarbons (PAHs), are responsible for the UV bump; small grains in general drive the steep FUV rise; and larger silicate grains dominate the optical-near infrared (NIR) \citep{Weingartner01,Draine03}. At the scale of molecular clouds and star-forming regions, the geometry between young stars and their birth clouds sets how much UV emission is attenuated before escaping: stars younger than $\sim 10\,\mathrm{Myr}$ remain embedded and experience higher attenuation than older populations that have dispersed into the diffuse ISM \citep{CharlotFall2000,Narayanan18,Matsumoto26}. On galaxy-wide scales, radiative-transfer (RT) effects -- scattering, multiple photon paths, and the spatial mixing of sources and dust -- systematically flatten attenuation curves relative to the underlying extinction curve \citep{Witt00,Chevallard13,SeonDraine16}. Hydrodynamical simulations post-processed with RT show that curves steepen when old stellar populations dominate and flatten when young stars are unobscured due to feedback-carved low-density channels \citep{Narayanan18,Trayford20}, and that vastly different extinction curves can produce similar attenuation features, underscoring the difficulty of inferring grain properties from integrated SEDs \citep{Lin21}. At smaller scales, feedback and turbulence within a single GMC can drive significant attenuation curve variation — including UV bump suppression — even at fixed dust composition \citep{DiMascia24}.

An additional complication is that dust composition and grain size distributions are themselves expected to evolve alongside galaxy assembly and ISM processing.
Live dust models — both analytic \citep{Asano13,Hirashita19,Parente26} and embedded in cosmological or zoom simulations \citep{Li_2019,Aoyama:2020,Choban2022,Choban2024,Dubois24,Narayanan25,Montero26} — predict that large grains, predominantly produced by stellar sources (AGB stars, supernovae), dominate the early ISM, while smaller grains are progressively built up through ISM processing (shattering in turbulent media, accretion onto pre-existing grains). This implies a coupling between the grain size distribution, metallicity, and the galaxy's star formation and chemical enrichment history, and predicts a tendency toward flatter, grayer extinction curves and weaker UV bumps at low metallicity and at high redshift — qualitatively consistent with the observational trends described above.

In this context, \citet{Dubois24} implemented a two-size, two-composition (carbonaceous and silicate) dust evolution model in the RAMSES code, and applied this to a suite of isolated disk simulations spanning a range of masses and metallicities. They found that the transition from ejecta-dominated to accretion-dominated grain growth occurs around $\simeq 0.1$--$0.5\,Z_\odot$ depending on grain composition \citep[see also][]{Asano13}, producing systematically steeper UV slopes and weaker $2175\,\angstrom$ bumps in lower-mass, lower-metallicity galaxies -- consistent with the observed trend toward SMC-like extinction at low metallicity. \citet{Matsumoto26} used isolated galaxy simulations to disentangle the contributions of individual physical effects, turning on and off scattering and comparing static versus live dust models. They find a multi-causal picture in which scattering -- and its dependence on column density along and across lines of sight -- emerges as the primary driver of attenuation curve shape, with grain size distribution and the unobscured-star fraction also substantially affecting the UV bump.

\textit{In summary}, the attenuation curve shape shows correlation with and/or is shaped by galaxy stellar mass, specific star formation rate (sSFR), metallicity, optical depth, stellar population age, the density and structure of birth clouds, and the grain size distribution and composition. Disentangling the roles of each of these physical quantities, which are themselves correlated, remains a central challenge in the field.

Despite this physical complexity, attenuation curves in both observational inference and forward modeling are typically compressed into empirically-inferred analytic parameterizations, and the choice of parameterization can substantially affect the inferred galaxy properties and any claimed evolutionary trends. 
Although the low-dimensionality of attenuation curves is generally accepted, it is not clear \textit{a priori} how many dimensions they need to have.
Several prescriptions are in common use, with a varying number of free parameters. The simplest is a \textit{power law}, $A_\lambda / \Av \propto (\lambda/5500\,\angstrom)^\delta$, which captures the overall slope with a single parameter. \citet{Calzetti00} provided a widely used piecewise polynomial calibrated on local starbursts. \citet{Noll2009} generalized the Calzetti law with a power-law tilt of slope index $\delta$ and a UV bump of variable amplitude $B$ modeled as a Drude profile, a parameterization later applied by \citet{Salim18} to a large sample of low-redshift galaxies, with an additional $R_V$ normalization (see \cref{eq:salim18}). \citet{CharlotFall2000} modeled the attenuation as the sum of two power-law components, representing birth clouds and the diffuse ISM, each with its own $V$-band normalization and slope. \citet{Li08} proposed a flexible analytic form capturing the FUV rise, the optical-NIR slope, and the UV bump independently via four parameters $c_1$--$c_4$ (see \cref{eq_Li08}).

The functional form adopted in SED fitting is not a neutral choice: the number of free parameters and choice of parameterization impose different priors on the allowed attenuation curve shapes, and can introduce biases on inferred galaxy parameters (such as stellar mass or star formation rate) or artificial degeneracies -- for example, between the UV bump amplitude and the overall slope, or between the attenuation curve shape and other galaxy parameters such as the star formation history. This may be a significant and underappreciated source of systematic uncertainty in studies of attenuation curve evolution.

\begin{figure*}   
    \includegraphics[width=\linewidth]{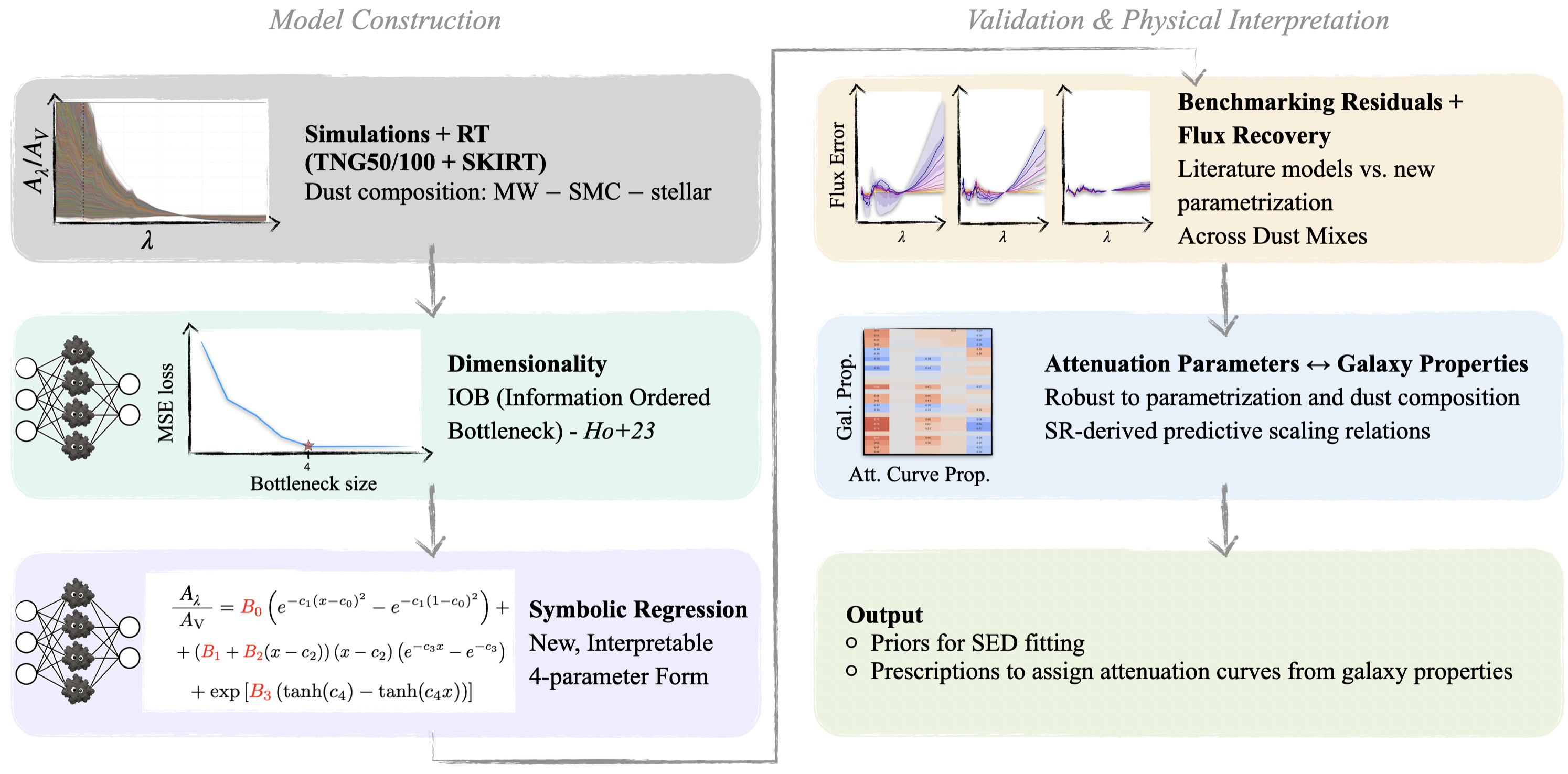}
    \caption{Schematic overview of the methodology and main results of this work.
    }
    \label{fig:scheme}
\end{figure*}

In a previous paper, \citet{Sommovigo_2025} adopted the four-parameter \citet{Li08} functional form to fit synthetic attenuation curves extracted from IllustrisTNG50 and IllustrisTNG100 simulations post-processed with the \skirt{} radiative transfer code \citep{CampsBaes2020} assuming a MW dust mixture \citep{Weingartner01}, for a sample of nearly $6400$ galaxies. That work derived probability distribution functions for the best-fitting attenuation curve parameters and explored their correlations with global galaxy properties.

In the present work, we go substantially further by revisiting the functional form itself. Rather than assuming a specific parameterization, we ask: \textit{how many parameters are genuinely needed to describe the diversity of attenuation curves in our simulated library}, and \textit{what functional form best captures attenuation curve diversity?} To answer this, we employ: (i) \textit{dimensionality reduction} to determine the intrinsic dimensionality of the attenuation curve space; (ii) \textit{symbolic regression} to discover interpretable functional forms; and (iii) a systematic \textit{comparison of the fitting performance} of our new expressions against previously used forms in the literature, crucially across different dust mixture models. Finally, we study the correlations of the best-fitting parameters -- across all functional forms -- with galaxy properties and among parameters, highlighting the key improvements enabled by our newly proposed parameterization.

This paper is part of the Simons Collaboration on ``Learning the Universe''\footnote{\url{http://learning-the-universe.org}} (LtU), which aims to infer the initial conditions and physical laws governing our Universe from galaxy observations using a Bayesian forward-modeling approach. Generating accurate synthetic observables within this framework requires a faithful and computationally tractable model for dust attenuation, a key ingredient shaping predicted galaxy SEDs. The attenuation model and scaling relations derived here are designed to meet this requirement, enabling realistic dust prescriptions to be assigned in forward models and during inference without requiring expensive radiative-transfer calculations.

The paper is organized as follows. In Section~\ref{sec:data} we describe the IllustrisTNG simulations and the \skirt{} post-processing procedure used to generate our synthetic attenuation curve library. Section~\ref{sec:iob} presents the dimensionality analysis. Section~\ref{sec:symreg} describes the symbolic regression approach and resulting functional forms, with our final model given by \cref{eq:final_model}. Section~\ref{sec:appl_form_TNG} compares the performance of different parameterizations, dust mixtures, and the impact on degeneracies among attenuation curve parameters. Section~\ref{sec:galaxy_correl} presents the correlations between attenuation curve parameters and galaxy properties. Section~\ref{sec:sr_galaxy_prop} presents a minimal set of scaling relations derived via symbolic regression, aimed at describing the attenuation curve best suited to a given galaxy within our new parameterization. We compare with the literature and discuss caveats of this work in Section~\ref{sec:discussion}, then draw conclusions in Section~\ref{sec:conclusions}. The structure of the paper is summarized schematically in \cref{fig:scheme}.

\section{Simulated attenuation curves \label{sec:data}}

The majority of our synthetic attenuation curves are drawn from the post-processed IllustrisTNG\footnote{\url{https://www.tng-project.org}} sample described by \cite{Sommovigo_2025}. The sample comprises $2399$ sources from the TNG50 \citep{Pillepich19} suite and $4887$ sources drawn from the TNG100 \citep{Pillepich18} suite, all at $z=0.07$ (snapshot 93)\footnote{We augmented the \citep{Sommovigo_2025} sample by 893 sources, 540 of which from TNG50 and the remaining from TNG100}. 
We include all TNG50 galaxies with non-negligible star formation rate (SFR) and metal content, while for TNG100 we impose a stellar mass cut of $M_{\star} > 10^{10}\,\rm M_{\odot}$ to ensure a well-populated massive end. 
The parameter space spanned by this galaxy sample is: $9 \leq \log_{10} (M_{\star}/\rm{M_{\odot}}) \leq 12.54$, $-4.19 \leq \log_{10} (\mathrm{SFR_{100Myr}/M_{\odot}yr^{-1}}) \leq 1.58$ in SFR averaged over $100\,$Myr, $-0.64 \leq \log_{10} (Z_{
\rm{gas}}/\rm{Z_{\odot}}) \leq 0.84$ in gas phase metallicity, and $0.17 <\mathrm{Age_{\star}/Gyr}< 10.97$ in stellar age. 

\indent We model the continuum emission from far-UV to far-infrared wavelengths using the radiative transfer code \skirt\footnote{\url{https://skirt.ugent.be/root/_home.html}} 
\citep{Camps2015,CampsBaes2020}, broadly following the methods described by \cite{Schulz20} and \cite{Popping2022_tng}. We briefly summarize the technical details here. The emission from star particles is modeled according to their ages and metallicities, using \citet{Bruzual2003} stellar population synthesis models. The IllustrisTNG simulation suite does not directly follow the dust abundance of gas cells. Hence, we assume a dust-to-gas mass ratio, $D$, for gas cells (below $75,000\,\rm{K}$) that scales linearly with the gas-phase metallicity:
\begin{equation}\label{eq_RR14}
    D = \frac{1}{163}\ \frac{Z}{\mathrm{Z_{\odot}}}\ ,
\end{equation}
based on the empirical relation derived by \cite{RemyRuyer14}. 

Our fiducial dust model comprises a mixture of graphite, silicate and PAH grains, with sizes in the range $a=10^{-4}-10\,\mu\rm{m}$ according to the \cite{Weingartner01} Milky Way dust model. 
In reality, dust grain composition varies from galaxy to galaxy and even within individual galaxies, 
depending on local ISM conditions, metallicity, and star-formation history. 
Rather than attempting to model this variation explicitly, which would require live dust 
evolution models beyond the scope of this work, we approximate the expected diversity in grain 
populations by running three independent radiative-transfer suites, each adopting a globally uniform 
but physically distinct dust composition. By combining these suites, our attenuation curve library samples 
a broad range of grain properties, from MW-like mixtures with significant PAH and small-grain fractions, 
to SMC-type and large-grain-dominated stellar dust.
This approach is necessarily an approximation, but it allows us to span the plausible range of dust mixture 
space in a controlled and computationally tractable way.
  
We perform the radiative transfer on a Voronoi dust grid, in which cell sizes are adjusted according to the dust density distribution, with the condition that no dust cell may contain more than $0.0001\%$ of the total dust mass of the galaxy. We calculate rest-frame far-UV to far-IR emission (at 200 wavelengths in the range $0.1-10^3 \, {\rm \mu m}$) along $51$ lines of sight, evenly distributed in solid angle. 
Although radiative transfer is computed for a wider range of wavelengths, in this work we only consider wavelengths below $1\,{\rm \mu m}$ since we are not attempting to model infrared thermal re-emission.
Comparing the ratio of the modeled observed emission along a given line of sight to the intrinsic (dust-free) emission, we calculate integrated dust attenuation curves along each line of sight. Following \cite{Popping2022_tng}, we do not model the contribution from young birth clouds.

We additionally perform radiative transfer calculations using the \cite{Weingartner01} SMC and the \cite{Hirashita19} stellar dust models (the latter described by a log-normal grain size distribution with centroid $a_0 = 0.1\,\mu$m and width $\sigma = 0.47$, in the range $0.001$--$10\,\mu$m), for a subset of galaxies sampling the full parameter space in $M_\star$ and SFR.
To select this subsample, we divide our initial sample of $\sim 7300$ TNG50 and TNG100 galaxies into 30 adaptive (equal-count quantile) bins in $\log_{10}(M_\star/M_{\odot})$, ensuring each bin contains a comparable number of galaxies. Within each stellar mass bin, we then select 15 galaxies uniformly spanning the $\Sigma_{\rm SFR}$ distribution, since \citet{Sommovigo_2025} found $\Sigma_{\rm SFR}$ to encode a significant fraction of the attenuation curve diversity in the MW dust case. 
This yields $\sim 450$ galaxies that systematically sample the $\log M_\star$–$\Sigma_{\rm SFR}$ plane.
To ensure adequate coverage of extreme attenuation regimes, we supplement this selection with 32 additional galaxies drawn from the tails of the $\Av$ distribution. The final subsample contains 482 galaxies (148 from TNG50 and 334 from TNG100), for which we perform radiative transfer calculations with SMC and stellar-type dust (on top of the already available MW dust). A posteriori, we verified that the subsample adequately reproduces the dimensionality and parameter space coverage of the full MW attenuation curve sample, with negligible changes in the recovered parameter percentiles.

\section{Intrinsic dimensionality of attenuation curves \label{sec:iob}}

From the library of synthetic dust attenuation curves computed in the previous section, we build a phenomenological model for dust attenuation as a function of emission wavelength to capture the variability in our dataset. The first step in constructing this model is determining the optimal number of free parameters required to describe the underlying variability within the dataset. 

For this, we infer the intrinsic dimensionality using an Information-Ordered Bottleneck (IOB; \citealt{Ho_2023}).
The IOB utilizes a specialized neural autoencoder to compress and reconstruct the spectral information contained in the dust attenuation curves. It does so by routing all information through a narrow intermediate layer (the bottleneck), whose width, defined as the number of active neurons, controls how much information is preserved. The key insight of the IOB autoencoder is that this bottleneck width is varied during training: individual neurons are randomly deactivated with a probability that increases with their index, enforcing a hierarchy where the first neurons capture the most important features and later ones progressively less. The width at which reconstruction quality stops improving then identifies the intrinsic dimensionality of the dataset, i.e., the minimum number of parameters needed to faithfully describe the diversity of attenuation curves in our library.
A related autoencoder approach has previously been used to determine the optimal number of free parameters to describe halo density profiles \citep{Lucie-Smith_2022,Lucie-Smith_2024}, although using a mutual-information-based loss, as opposed to the IOB loss we adopt here.

We train the IOB using a symmetric autoencoder neural network with a depth of 6 layers and a maximum bottleneck width of 8. We train it to adaptively compress and reconstruct the 237,588 dust attenuation curves from the parent MW dust run, using a mean-squared-error (MSE) likelihood with additional penalty terms on the bottleneck activations: an $\ell_2$ penalty to prevent any single activation from dominating, and a covariance penalty to ensure each bottleneck dimension captures independent information. We train until convergence using an Adam optimizer \citep{Kingma_2014}, stopping when validation performance shows no improvement over 10 epochs. These hyperparameters are set by manual tuning, as automated searches have not been validated in the IOB context and could add uncertainty to the dimensionality estimates.  We find that at full bottleneck width, we can reconstruct  99.95\% of the variance of our original dust models. We note that the same intrinsic dimensionality is recovered when the MW, SMC, and stellar dust attenuation curves are analyzed jointly.

\begin{figure}
    \centering
    \includegraphics[width=\columnwidth]{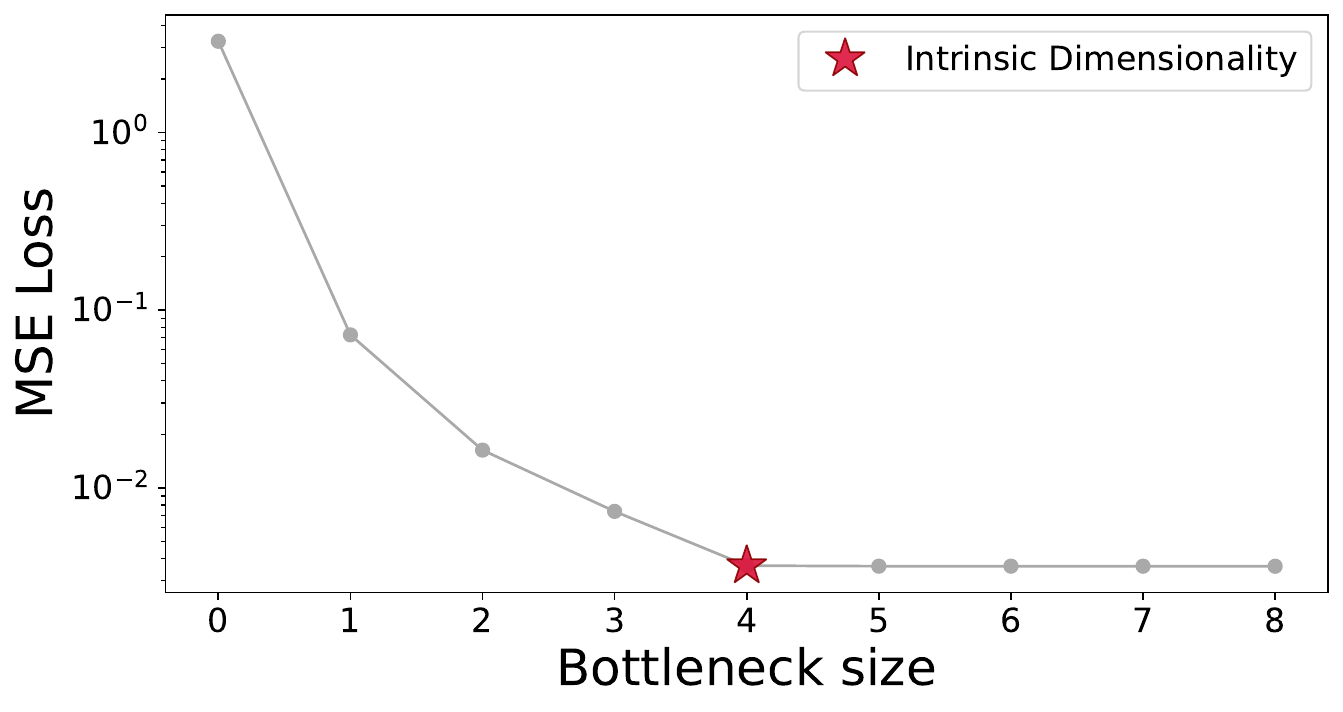}
    \caption{Reconstruction MSE as a function of bottleneck size for the MW dust synthetic attenuation-curve test set. The red star marks the inferred intrinsic dimensionality of the dataset. We recover the same intrinsic dimensionality when the MW, SMC, and stellar dust attenuation curves are analyzed jointly, indicating that the full diversity of TNG attenuation curves can be faithfully captured by a four-dimensional representation.
    }
    \label{fig:iob_dim}
\end{figure}

\cref{fig:iob_dim} shows the average reconstruction performance of our dust attenuation dataset as a function of the bottleneck width of the autoencoder. The reconstruction error converges at four open nodes, which corresponds to an intrinsic dimensionality of four based on the convergence tests established in \citet{Ho_2023};
we perform a likelihood ratio test: the bottleneck size beyond which adding an extra latent dimension no longer yields a statistically significant improvement in reconstruction accuracy (at the $p = 0.05$ level, compared to a $\chi^2$ distribution with one degree of freedom).
The trained network functionally serves as a phenomenological model for dust attenuation, conditioned on the four latent variables passed through the trained decoder architecture. In the next section, we utilize this information to derive a simpler, more expressive symbolic expression to replace the learned neural representation. 

\section{A new functional form for attenuation curves \label{sec:symreg}}

To obtain a new functional form for the attenuation curve, we use the supervised machine learning technique of symbolic regression \citep[SR; see][for a recent review]{Kronberger_2024}.
We wish to obtain a single function of wavelength, $\lambda$, which can fit all attenuation curves, with some free parameters which can be different for different curves (i.e. different galaxies, lines-of-sight, and dust mixtures).
Within SR, one can achieve this by searching for a common structure of the equation and explicitly fitting different parameters to each distinct subset of the data \citep{Tenachi_2024,Russeil_2024,Russeil_2025}. For very large training datasets, this involves optimizing a large number of parameters, which is computationally expensive. As an alternative, in \citet{Martin_2025,Martin_2026}, each subset of the data is analyzed completely independently, and in post-processing one checks whether certain parameters can be made `global', i.e. are the same for all objects.

However, we approach this problem differently by splitting our problem into two parts. 
In the preceding section, we found that we could express the attenuation curve $A_\lambda/\Av$ in terms of wavelength, $\lambda$, and four other parameters ($I_1$, $I_2$, $I_3$ and $I_4$), which are the latent variables of the IOB when the bottleneck size is set to four.
Given this embedding, the attenuation curve was parameterized as a neural network (the decoder) in Section~\ref{sec:iob}.
To give a more interpretable result, in this section we find a compact analytic expression for the decoder of the network using symbolic regression (SR).
The latent variables then become the free parameters of the function, which can be optimized when applied to a previously unseen attenuation curve.
Throughout this section, we derive this function using the MW dust mixture (and thus the IOB embedding relevant to these attenuation curves). 
This results in the functional form given by \cref{eq:final_model}, which is the fitting function which we advocate for in this work.
In Sec.~\ref{sec:appl_form_TNG}, we test how our model extrapolates to other dust mixtures.

\subsection{Symbolic regression}

\begin{figure*}
    \includegraphics[width=\textwidth]{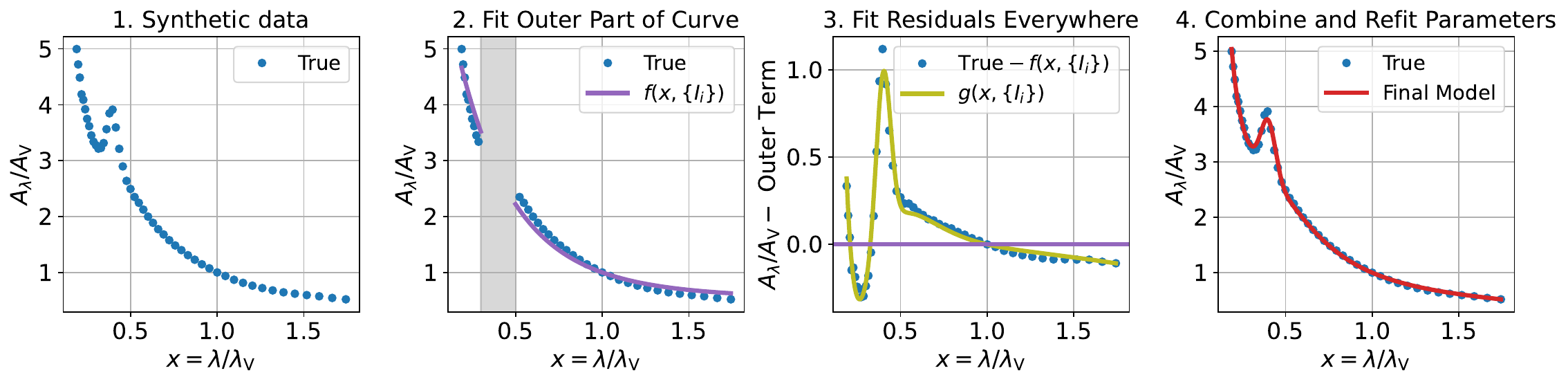}
    \caption{Procedure for obtaining a functional form for the attenuation curve, $A_\lambda / \Av$, as a function of wavelength, $\lambda$, and the IOB latent variables, $\{I_i\}$, with symbolic regression. We find that more interpretable results are obtained if we first fit a function to the broad-band features of the curve, $f(x, \{I_i\})$, by masking the wavelength range containing the bump (gray band in second plot). An analytic expression for the residuals of this function is then found, $g(x, \{I_i\})$, before the terms are summed and any free parameters are re-optimized.
    \label{fig:fitting_procedure_sr}
    }
\end{figure*}

We make use of the genetic programming based SR code \operon\footnote{\url{https://github.com/heal-research/pyoperon}} \citep{Burlacu_2020}, because of its speed, memory efficiency, strong performance on benchmarks \citep{LaCava_2021,Burlacu_2023} and its established effectiveness in cosmological and astrophysical research \citep[e.g.][]{Bartlett_2024_syren,Bartlett_2024_linear,AbdusSalam_2025,AbdusSalam_2024,Bartlett_2025,Farakou_2025,Kammerer_2025,Sui_2024}.
\operon{} evolves a `population' of functions over many generations in a natural-selection based approach; functions are allowed to exchange parts of their expressions or `mutate' some of their operators to produce new individuals. The free parameters of these functions are optimized using the Levenberg–Marquardt algorithm \citep{Levenberg_1944,Marquardt_1963} as described in \citet{Kommenda_2020}, and those with the worst loss values are removed at each iteration. Over successive generations, the population `evolves' to better-fit the training dataset.
We make use of a dual-optimization strategy, where we jointly minimize the error on the training data and the model length (approximately equal to the number of symbols appearing in the expression).

As is usual with SR, we have the freedom to manipulate the search for a reasonable symbolic expression because we need to find at least one good, interpretable one, instead of a theoretical global MSE-minimizing optimum.
Hence, rather than directly finding a single function to fit the full attenuation curve, we found that more interpretable results were obtained if we split the problem into two parts
\begin{equation}
    \frac{A_\lambda}{\Av} = f(x, \{I_i\}) + g(x, \{I_i\}),
\end{equation}
where $x \equiv \lambda / \lamv$ for $\lamv = 5542 \, \angstrom $, and $f$ is chosen to fit the `outer' part of the attenuation curve away from the bump feature, whereas $g$ is designed to fit the region containing the bump.
This procedure is summarized in \cref{fig:fitting_procedure_sr}.
This makes an implicit assumption that we can interpret the inner and outer trends of the attenuation curve separately.
Although here we explicitly write the dependence on $\{I_i\}$ as an argument of $f$ and $g$, for brevity, in the remainder of the paper we drop this and simply write $f(x)$ and $g(x)$.

\subsection{Data subsampling}

In Section~\ref{sec:data}, we described how we obtained attenuation curves by post-processing galaxies from a hydrodynamical simulation. For our SR fits, we do not wish to use all of these curves, but merely a representative subset that are designed to span stellar mass, star-formation surface density ($\Sigma_{\rm SFR}$), and dust attenuation in a controlled and balanced way.

To obtain these, we began by removing 0.2\% of all attenuation curves for which $A_\lambda/\Av$ had either negative or non-finite values. Negative attenuation values can occur when photons scattered from more heavily obscured directions boost the received flux above the intrinsic value along nearly dust-free lines of sight, while non-finite values arise when the V-band attenuation is negligibly small, rendering the normalisation $A_\lambda/\Av$ ill-defined. 
We also removed some pathological cases showing extremely peaked curves, for which $A_\lambda/\Av$ contained very sharp jumps at long wavelengths, likely associated with numerical artifacts due to limited photon statistics in the radiative-transfer calculation.
This was done by considering wavelengths $\lambda \geq \lamv$ and removing attenuation curves for which $A_\lambda/\Av$ changed by more than $0.5$ between two successive sampled wavelengths.
This is only a very small fraction of our sample, consisting of just 0.4\% of all attenuation curves.

Since we wish to consider attenuation curves with a diverse range of galaxy properties, we next partitioned our galaxies into 30 adaptive stellar mass bins such that each contained approximately the same number of galaxies. 
We chose the same number of galaxies from each of these bins at equally-spaced percentiles in $\Sigma_{\rm SFR}$.
We explicitly augmented this sample with rare or extreme galaxies that would otherwise be underrepresented. These are defined as those with $\Av$ values above the 84\textsuperscript{th} percentile and with $\Sigma_{\rm SFR}$ above $0.1 \, \sfrunit$ or those with $\Av$ below the 20\textsuperscript{th} percentile and with $\Sigma_{\rm SFR}$ below $10^{-3} \, \sfrunit$.
5\% of our sample was chosen to comprise galaxies with extremely high values of $\Av$ and $\Sigma_{\rm SFR}$, and 1.2\% of the sample contains the extreme low values.
The remainder of each sample was then drawn according to the procedure outlined above.

We repeated this procedure to build a disjoint training and validation set, containing 1500 and 750 attenuation curves, respectively.

\subsection{Learned analytic expression for outer region}

\begin{figure*}[htb]
  \centering
  \includegraphics[width=\textwidth]{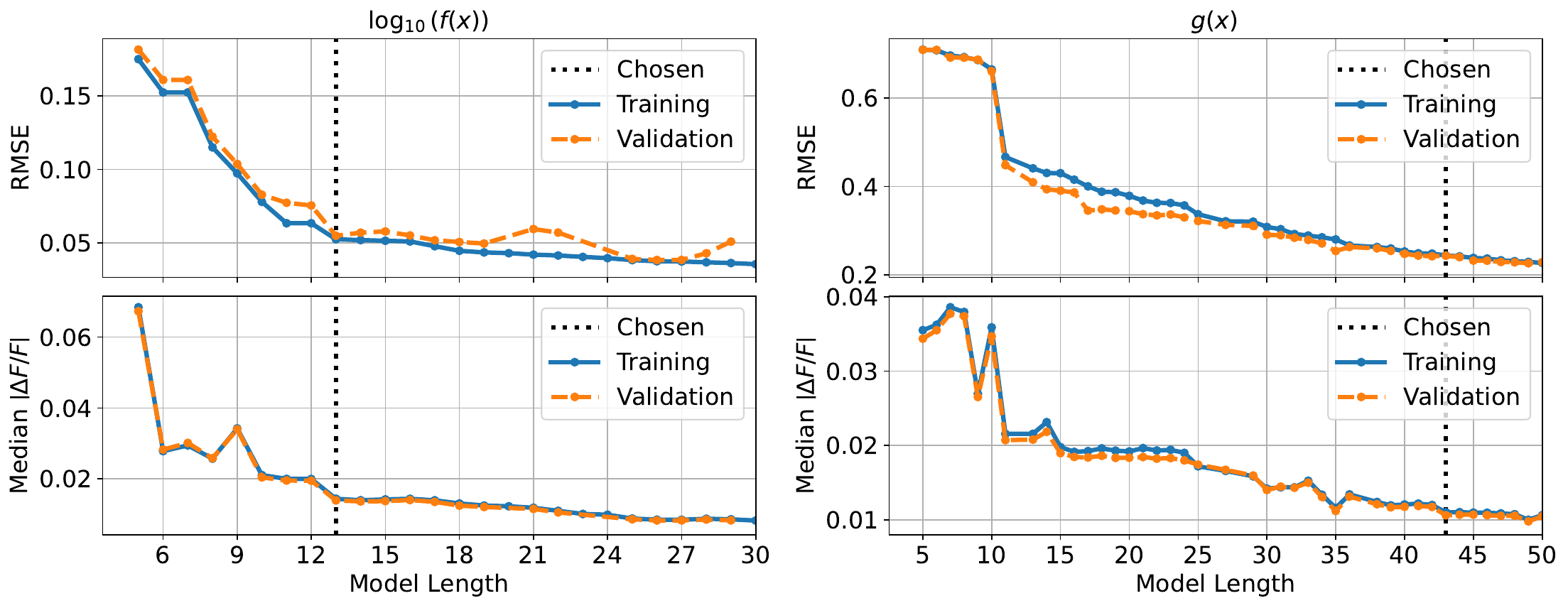}
  \caption{Pareto front (equations minimizing RMSE for a given model length) of solutions found with \operon{} for (left) the outer part of the attenuation curve, $\log_{10}(f(x))$, and (right) the bump region of the attenuation curve, $g(x)$. On the top, we plot the root mean squared error (RMSE) on the target variable, and on the bottom we plot the corresponding median absolute fractional error on the corresponding flux, $\Delta F / F$. We plot the training and validation sets separately, and the vertical lines indicate our chosen models.
  }
  \label{fig:pareto}
\end{figure*}

To obtain an equation for the longer wavelength regime of the attenuation curve, $f(x)$, we first removed all points in the wavelength range $0.166 - 0.277 \, {\rm \mu m}$.
Since the attenuation curve has a reasonably large dynamic range between the smallest and largest values of $x$, and given that these are the regions we are attempting to fit, we found it appropriate to fit for $\log_{10} f$ instead of $f$ to bring these two regimes onto the same scale.
We searched for functions of a maximum length of 30 comprising standard arithmetic operators ($\splitatcommas{+, -, \times, \div}$) as well as the square, square root, logarithm, power and hyperbolic tangent operators.
We ran \operon{} for 4 hours on 28 CPU cores, jointly minimizing the root mean squared error between the true and predicted $\log_{10}(A_\lambda/\Av)$ and the model length.

This resulted in \cref{fig:pareto}, where we plot the minimum error found at each model length (also called the `Pareto front') and indicate our chosen model (of length 13) with a dotted vertical line.
We see that, beyond the chosen model length, both the root mean squared error on $\log_{10}(f(x))$ and the median absolute error of the fractional difference in the flux $\Delta F/F$ plateau, indicating negligible improvement for more complex models.
We visually inspected several of the functions along the Pareto front to make an aesthetic judgment of the best function which balanced both accuracy and simplicity.

Since the IOB parameters are only used as a tool for reducing the dimensionality of the attenuation curve and do not necessarily carry any physical meaning, we merged all functions of these parameters into new parameters which can be optimized. After doing this and merging superfluous constants, our chosen model for the outer part of the attenuation curve is
\begin{equation}
    \label{eq:initial_outer_term}
    \log_{10} f(x) \approx C_0 \left( a_2 - \tanh (a_1 x) \right) + a_0,
\end{equation}
where $C_0$ is a function of $I_1$, and $\{a_i\}$ are constants.
Throughout, we use the convention that, for SR-derived models, lower-case letters refer to constants (which are the same for all attenuation curves), and upper-case letters refer to free parameters, which can take different values for each attenuation curve. For functional forms taken from the literature, we do not use this convention and use the standard notation used for those functions.

By definition, we want to enforce that $A_\lambda/\Av = 1$ at $\lambda = \lamv$ for all curves, which means that $f(1)$ must be independent of $C_0$.
This can be achieved if $a_2 = \tanh(a_1)$, which we found to be approximately true for the optimized values found by \operon. We therefore enforced this exactly to obtain
\begin{equation}
    \label{eq:final_outer_term}
    f(x) = b_0 \exp \left[ D_0 \left( \tanh(b_1) - \tanh(b_1 x) \right) \right],
\end{equation}
where we redefined the constants $a_0$ and $a_1$, and the free parameter $C_0$ to obtain the new constants $b_0$ and $b_1$, and the new free parameter $D_0$.

Since we have replaced a function of the IOB parameters with a new free parameter, we chose to re-optimize $D_0$ since one can improve the loss with this added freedom. This also changes the optimal values of $\{b_i\}$, so we approached this in a two-stage optimization. We began by fixing $D_0$ to its value obtained from the IOB parameters and re-optimized the constants $\{b_i\}$ by minimizing the root mean squared error on the resulting fractional error on the flux, $\Delta F/F$, for the training set. We then fixed the newly optimized $\{b_i\}$, and we found a new optimal $D_0$ for each attenuation curve in both the training and validation sets by minimizing the mean squared error on $A_\lambda/\Av$ in the region excluding the bump. 
We re-optimized all parameters after finding the full attenuation curve, so the purpose of this optimization is to obtain a reasonable guess of $f(x)$ to aid with our search for $g(x)$.
Since the values of $\{b_i\}$ and $D_0$ are re-optimized later, we do not state their values here.

\subsection{Learned analytic expression for bump region}

Now that we have an optimized term for the outer part of the attenuation curve, we wish to find an analytic expression for the region containing the bump. 
Using the same training dataset as before, but now using the full range of wavelengths, we subtracted the prediction for the outer part of the curve, and fitted these residuals using \operon.
For this term, we found that we obtained better results by using a different set of basis operators to before, by removing the logarithm, power and hyperbolic tangent operators, and adding in the exponential operator.
Since we expect this term to be more complex than the `outer' term to well-capture the bump feature and any residuals from the expression in \cref{eq:final_outer_term}, we increased the maximum model length to 50 and increased the run time to 12 hours on 28 cores.

The resulting best loss at each model length is given in \cref{fig:pareto}, where our chosen model of length 43 is again indicated with a dotted vertical line.
This function was chosen since there are diminishing returns with regards to accuracy by going to longer model lengths, and this function is sufficiently accurate to achieve approximately a 1\% median fractional error on the flux. 
We inspected several models from the \cref{fig:pareto} and from many repeats of the fitting with various random seeds to ensure we obtained the simplest model possible while remaining highly accurate.

After merging functions of IOB parameters into new parameters and simplifying, our chosen function can be written as
\begin{equation}
    g(x) \approx a_0
    + C_0 e^{-a_2(x-a_1)^2}
    + (a_3 + x)(C_1 + C_2 x) e^{a_4 x},
\end{equation}
where we note that the parameters $\{a_i\}$ and $\{C_i\}$ are different to those given in \cref{eq:initial_outer_term}.

Unlike for the outer term, it is not possible to round the optimized constants of this function such that $A_\lambda/\Av$ is independent of $\{C_i\}$ at $x=1$.
However, using the optimized $a_1$ and $a_2$, we found that the Gaussian ($e^{-a_2(x-a_1)^2}$) multiplying $C_0$ is approximately $e^{-100}$ at $x=1$, making this term negligible. Nevertheless, to explicitly enforce this to be zero, we replaced the Gaussian with a Gaussian minus its value at $x=1$.
When considering the final term for $g(x)$, we found that $e^{b_4 x}$ is approximately $10^{-4}$ at $x=1$, so again this makes this term very small. In a similar way to the Gaussian term, we replaced $e^{a_4 x}$ with $e^{a_4 x} - e^{a_4}$ to enforce that this term vanishes at $x=1$. After making these two substitutions, one sees that $g(x=1)=a_0$, so is independent of $\{C_i\}$, as required.
In fact, we found that the value of $a_0$ returned by \operon{} is approximately $-2\times10^{-3}$, so we rounded this parameter to zero.
To ensure the sum of $f(x)$ and $g(x)$ is unity at $x=1$, we therefore require $b_0=1$ in \cref{eq:final_outer_term}.

\begin{figure*}[htb]
  \centering
  \includegraphics[width=0.9\textwidth]{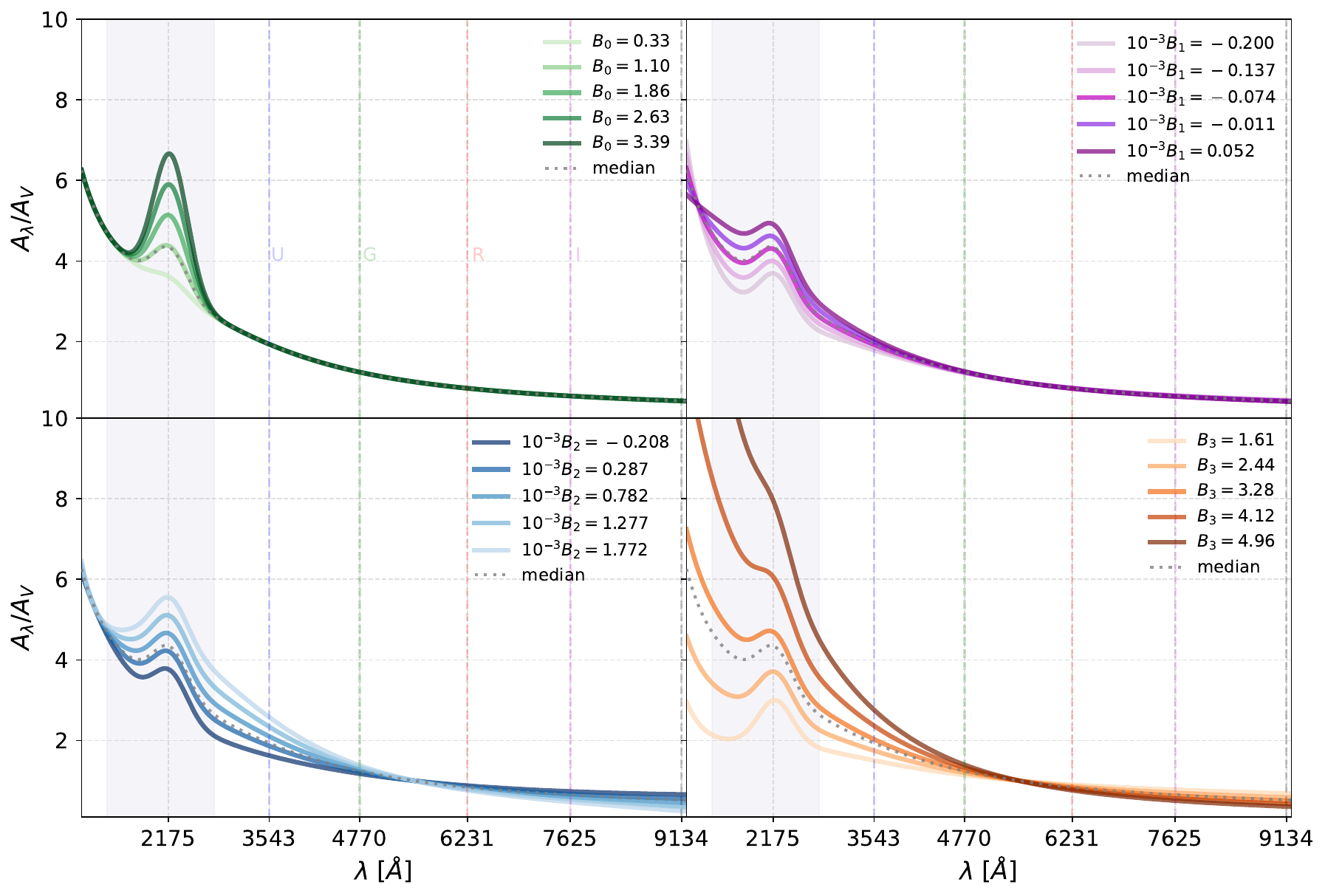}
  \caption{Summary plot illustrating the effect of independently varying each parameter of our new attenuation-curve model (\(B_0\), \(B_1\), \(B_2\), and \(B_3\); Eq.~\ref{eq:final_model}), while keeping the remaining parameters fixed at their median values. The grey dotted curve shows the median attenuation curve of the combined TNG galaxy sample, while the colored curves illustrate the impact of varying a single parameter across its central percentile range. The parameters primarily regulate the UV bump strength (\(B_0\); top left), the FUV slope (\(B_1\); top right), the curvature of the UV-to-optical transition region (\(B_2\); bottom left), and the large-scale UV-to-optical slope (\(B_3\); bottom right). The vertical colored lines indicate the central wavelengths of the SDSS filters (\(u\), \(g\), \(r\), \(i\), \(z\)), while the grey shaded region highlights the wavelength coverage of the GALEX bands.
  }
  \label{fig:param_example}
\end{figure*}

\subsection{Final learned analytic expression}

Having performed these manipulations, we sum our results for $f(x)$ and $g(x)$ and redefine our parameters and constants to obtain our final learned analytic expression for the attenuation curve
\begin{equation}
    \label{eq:final_model}
    \begin{split}
        \frac{A_\lambda}{\Av} &=
        B_0 \left[ e^{-c_1 (x - c_0)^2} - e^{-c_1 (1 - c_0)^2} \right] \\
        & + \left[ B_1 + B_2 (x - c_2) \right] \left(x - c_2 \right) \left( e^{-c_3 x} - e^{-c_3}\right) \\
        & + \exp \left[ B_3 \left( \tanh(c_4) - \tanh(c_4 x) \right) \right].
    \end{split}
\end{equation}
In particular, we note that we changed the term $C_1 + C_2 x$ to $B_1 + B_2 (x - c_2)$ since we found that this reduced the correlation between the free parameters of the function. We also redefined the constants so that they are all positive.

To obtain the optimized values of the constants $\{c_i\}$, we re-fitted both these constants and the parameters $\{B_i\}$ (giving four additional parameters to optimize for each attenuation curve) to our training dataset.
We used the Adam optimizer to minimize the mean squared error between the true and predicted values of $A_\lambda/\Av$. We used an initial learning rate of $10^{-2}$, and we reduced the learning rate by a factor of 2 if our loss did not improve for 10 epochs. We found that our parameters had converged within 1000 epochs, giving the final values
$\{c_i\} = \{ \splitatcommas{0.4002, 285.6, 0.2092, 9.223, 1.016} \}$.

The functional form can be interpreted by examining the roles of its individual terms. To visualize this, in \cref{fig:param_example} we plot the effect of varying each of the four parameters of \cref{eq:final_model} while keeping the others fixed at their median values (see Figure 3 of \citet{Sommovigo_2025} for an analogous plot for the \citet{Li08} model). 
The first line of \cref{eq:final_model} is a Gaussian term that characterizes the localized bump feature. 
Its amplitude, $B_0$, controls the strength of the bump and we therefore require it to be non-negative. 
In the absence of a bump, the functional form smoothly reduces to a bump-free attenuation curve when $B_0 = 0$. 
The Gaussian is centered at $c_0 \simeq 0.4$, corresponding to the location where the bump peaks, as shown in \cref{fig:fitting_procedure_sr}. 
Indeed, this bump is known to occur at $2175 \, \angstrom$ \citep{Cardelli89}, which is at $x \simeq 0.39 \approx c_0$.
Its width is $\sigma = 1/\sqrt{2c_2} \simeq 0.04$, which closely matches the characteristic width of the feature in \cref{fig:fitting_procedure_sr}. Notably, the mean and width of the Gaussian are the same for all attenuation curves, with only the amplitude left as a free parameter. This behavior was not enforced \textit{a priori}, but is a feature that SR learned. 

The behavior of the attenuation curve away from the bump is governed by an additional term containing the parameter $B_3$. This term asymptotically approaches a constant value at long wavelengths, and therefore $B_3$ controls the long-wavelength normalization of the attenuation curves, as well as the FUV rise. Since it represents a decaying component, we require $B_3 \geq 0$.

A further corrective term accounts for deviations at short wavelengths introduced by the exponential $\tanh$ dependence (which for small wavelengths can be approximated as an exponential). For fixed values of $B_0$ and $B_3$, the parameters $B_1$ and $B_2$ adjust the limiting value of the attenuation curve as $x \to 0$ as well as its slope and curvature in this regime. 
Together, these terms provide sufficient flexibility to reproduce the small-$x$ (short wavelength) behavior while preserving the physically motivated interpretation of the bump feature.
Due to being multiplied by a decaying exponential, these parameters have little effect on the attenuation curve at long wavelengths.
We can interpret this corrective term physically as encoding the contribution of the large-grain population to the FUV attenuation: at these wavelengths, larger dust grains become comparable in size to the impinging photons and approach the geometric-optics regime, where their absorption efficiency saturates and becomes only weakly wavelength-dependent \citep[e.g.][]{Draine03}. This contrasts with the small-grain population, which remains in the Rayleigh regime ($Q_{\rm abs} \propto \lambda^{-1}$) and continues to absorb increasingly efficiently toward shorter wavelengths \citep{Weingartner01}; the small grains therefore primarily shape the bump amplitude $B_0$ and the steeper FUV rise captured by $B_3$, while $B_1$ and $B_2$ provide the additional flexibility needed to accommodate the comparatively weakly varying large-grain contribution.

As will be discussed further in the following section, we find that both $B_0$ and $B_3$ take values that are typically $\mathcal{O}(1)$, except for bump-less curves, for which $B_0\approx0$. The dynamic range of $B_1$ and $B_2$ is larger, with $|B_1|$ typically up to a few hundred, and $|B_2|$ up to $10^3$. As such, for inference problems, we recommend fitting for $B_{1, {\rm s}} \equiv 10^{-3}B_1$ and  $B_{2, {\rm s}} \equiv 10^{-3}B_2$ so that all inferred parameters are close to unity.

\begin{figure*}
    \centering
    \includegraphics[width=1.\linewidth]{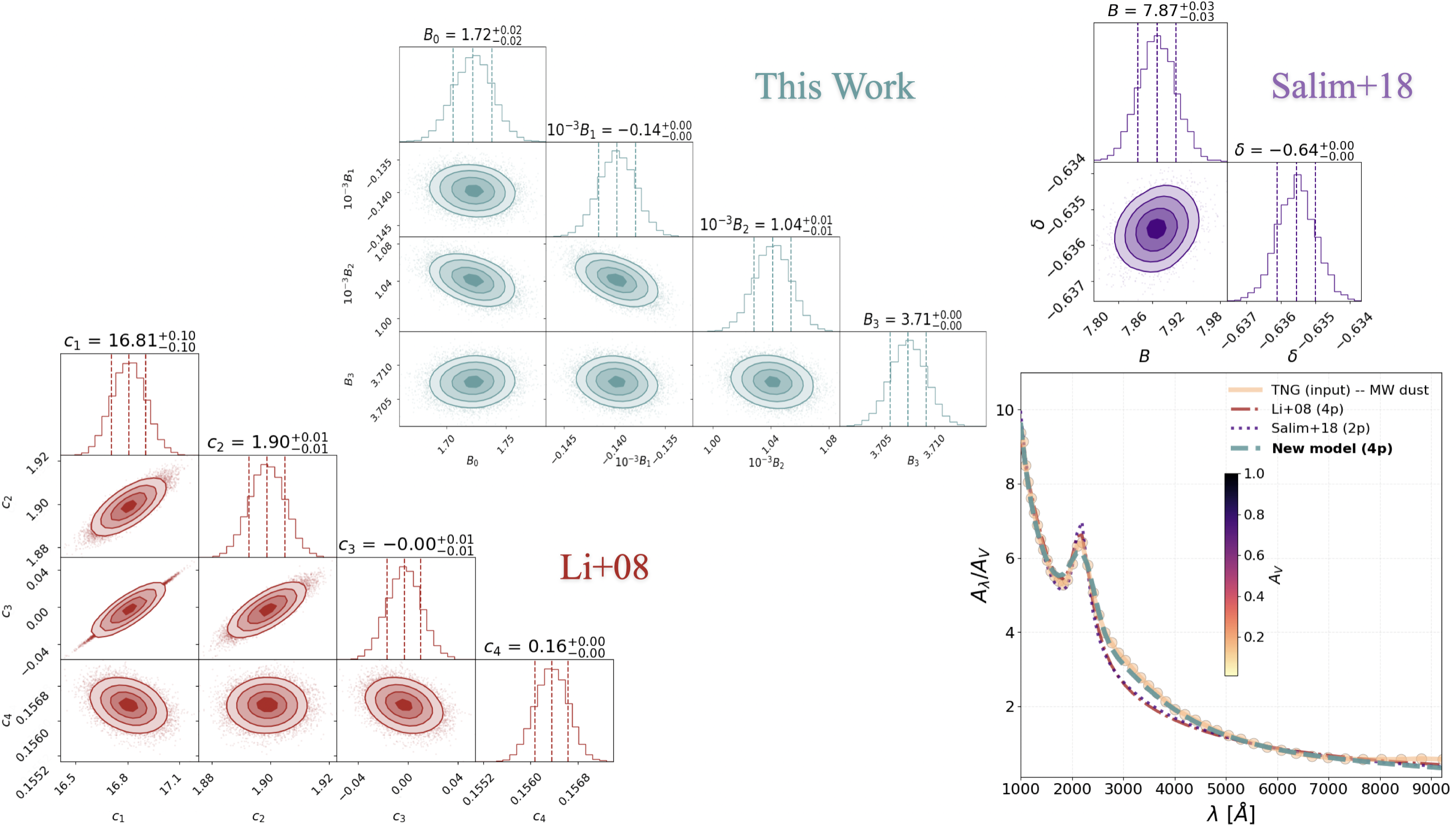}
    \caption{Posterior distributions of the best-fit parameters obtained by fitting the attenuation curve of a reference TNG50 galaxy (id~42) along a randomly selected line of sight (los~27). We show results using the 4-parameter model of \citet{Li08} (red), the 2-parameter model of \citet{Salim18} (purple), and our new symbolic-regression-based functional form (teal). The bottom right panel shows the input synthetic attenuation curve from the simulation (orange solid line) against the best-fit curve obtained with each parameterization in the respective color.
    Our model fits the attenuation curve the best, and the posterior distribution of the inferred parameters is less correlated than that of the \citet{Li08} model.
    }
    \label{fig:EMCEE_example_id42_los27}
\end{figure*}

\section{Application to synthetic attenuation curves from local galaxies}\label{sec:appl_form_TNG}

We now assess the performance of our newly derived functional form -- and those commonly used in the literature -- in reproducing the diversity of attenuation curves obtained from the radiative transfer (RT) post-processing of TNG galaxies. We focus first on the full sample of synthetic attenuation curves computed assuming a MW dust mixture, which constitutes both the largest dataset and the regime in which the model was derived. 
The dataset comprises a total of $237{,}588$ synthetic attenuation curves. Approximately $0.3\%$ of the initial sample is discarded due to numerical artifacts associated with negligible dust columns along the line of sight (see Section~\ref{sec:data}).

Each attenuation curve $A_\lambda / \Av$ is fitted independently for every galaxy and line of sight using three functional forms:

\begin{enumerate}
    \item the four-parameter \citet{Li08} parameterization ($c_1$, $c_2$, $c_3$, $c_4$):
    \begin{equation}\label{eq_Li08}
    \begin{split}   
& \frac{A_{\lambda}}{\Av} = \frac{c_1}{(\lambda/0.08)^{c_2}+(0.08/\lambda)^{c_2} +c_3} \\ 
& +  \frac{233[1-c_1/(0.145^{-c_2}+0.145^{c_2}+c_3)-c_4/4.60]}{(\lambda/0.046)^2+(0.046/\lambda)^2+90} \\ 
& + \frac{c_4}{(\lambda/0.2175)^2+(0.2175/\lambda)^2-1.95}\ ;
\end{split}
    \end{equation}
      
    \item the two-parameter \citet{Noll2009,Salim18} modification of \citet{Calzetti00}, introducing a UV bump amplitude $B$ and a slope deviation $\delta$:
    \begin{equation}\label{eq:salim18}
    \frac{A_\lambda}{\Av} = \frac{k_\lambda}{R_V} \left(\frac{\lambda}{\lamv}\right)^{\!\delta} + \frac{D_\lambda(B)}{R_V}\,,
    \end{equation}
    where $k_\lambda$ is the \citet{Calzetti00} curve, $R_V = 4.05$, and $D_\lambda(B)$ is a Drude profile with fixed $\lambda_0 = 2175\,\angstrom$ and $\gamma = 350\,\angstrom$;
    
    \item our new empirical four-parameter model ($B_0$, $B_1$, $B_2$, $B_3$; see \cref{eq:final_model}).
\end{enumerate}

For the new model, the sampled parameters are rescaled as $B_{1,s} = 10^{-3}\,B_1$ and $B_{2,s} = 10^{-3}\,B_2$ to improve numerical stability.

The fitting is performed in three stages. First, an initial least-squares estimate is obtained using \textsc{scipy.optimize.curve\_fit}. This solution is refined via minimization of the negative log-likelihood, and the posterior distribution is then sampled using the affine-invariant ensemble sampler \textsc{EMCEE} \citep{emcee}, assuming a Gaussian likelihood with $\sigma=1\%$ and uniform priors. We use $N_{\rm walkers}=32$ walkers initialized as small Gaussian perturbations around the maximum-likelihood solution, and run the sampler for 8,000 steps for the four-parameter models, and 4,000 steps for the two-parameter model (convergence was verified by increasing the number of steps).

The priors are chosen to be broad. For the \citet{Li08} model, we impose uniform bounds $c_1 \in (-10^3, 10^3)$, $c_2 \in (0, 100)$, $c_3 \in (-10^3, 10^3)$, and $c_4 \in (0, 0.8)$. For the \citet{Salim18} model, we adopt uniform priors $B \in (0, 100)$ and $\delta \in (-100, 100)$. For the new model, we require $B_0 > 0$ and $B_3 > 0$, leaving the remaining parameters effectively unconstrained.

We begin by establishing the relative performance of these models in the MW case. We then test the robustness of our results across different dust compositions, and finally examine the role of parameter degeneracies to assess whether the improved performance of our model arises from increased flexibility due to a relatively large number of parameters or from a more efficient representation of attenuation curve space.

\begin{figure*}[t]
    \centering
    \includegraphics[width=\linewidth]{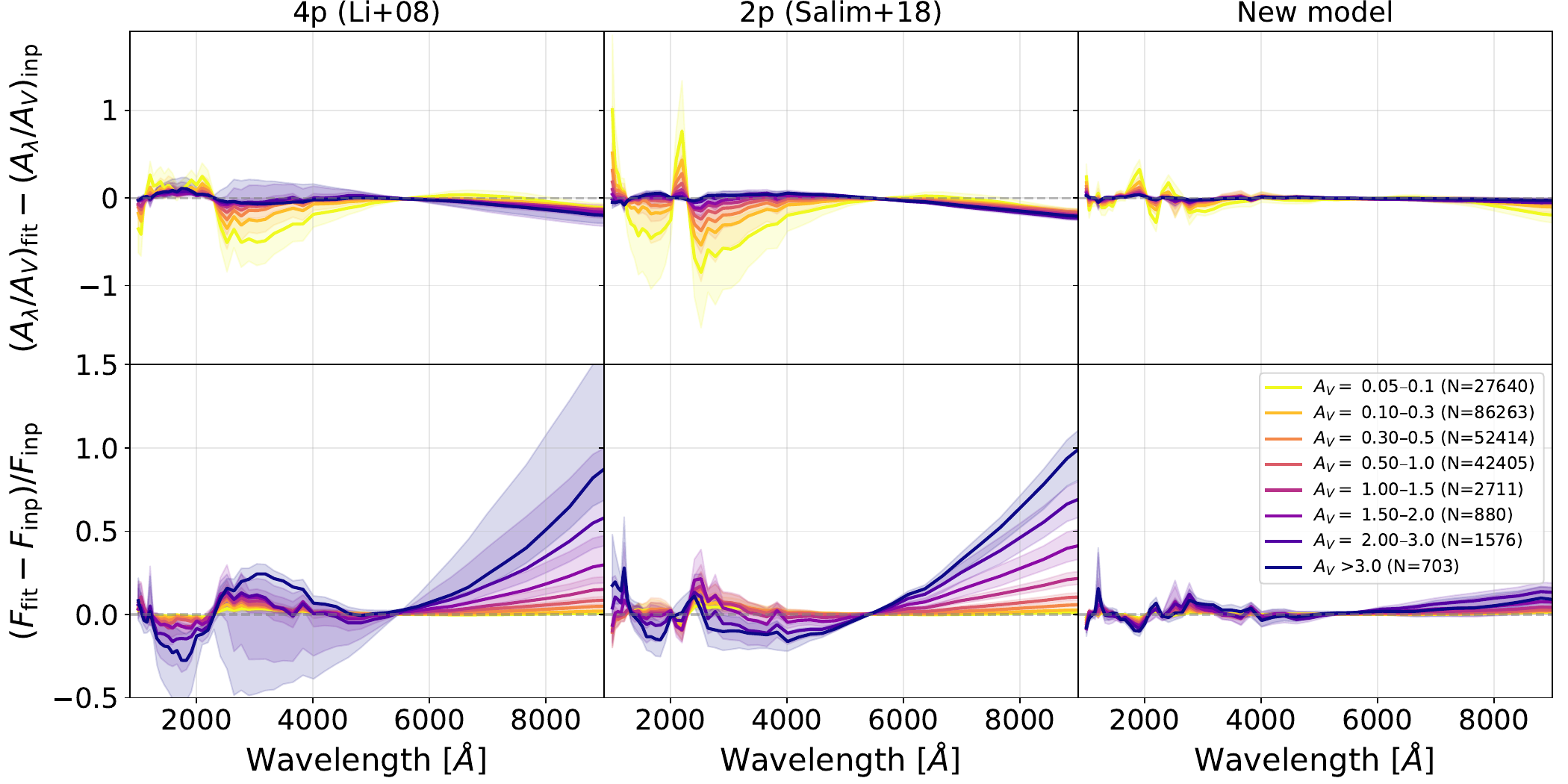}
    \caption{ Residuals in the normalized attenuation curve (top row) and fractional flux error (bottom row) for the three parameterizations: \citet{Li08} (4 parameters), \citet{Salim18} (2 parameters), and the new model (shown left to right), all adopting the fiducial \textbf{MW dust mixture}. For each galaxy and line of sight, the deviation between the best-fit and input attenuation curve is computed individually; the results are then binned by the line-of-sight $\Av$. Solid lines show the median within each $\Av$ bin, and shaded bands span the 30th–70th percentile range. The number of (galaxy, line-of-sight) pairs in each bin is indicated in the legend. 
    }
\label{Fig:summary_plot_EMCEE_MW}
\end{figure*}
    
\subsection{Performance for MW dust across parameterizations}

We now systematically assess the performance of the different functional forms in recovering both the attenuation curve and the observed flux for each line of sight processed with \skirt. An example with the three functional forms for a single randomly selected line of sight and TNG50 galaxy (id=42, also highlighted in \citealt{Sommovigo_2025}) is shown in Fig.~\ref{fig:EMCEE_example_id42_los27}.

For each synthetic attenuation curve, such as the one shown in Fig.~\ref{fig:EMCEE_example_id42_los27}, we compare the input $A_\lambda/\Av$ to the posterior median recovered from the three parameterizations and compute the residual $(A_\lambda/\Av)_{\rm fit} - (A_\lambda/\Av)_{\rm inp}$ as a function of wavelength. In addition, we evaluate the impact on the reconstructed flux by applying the best-fit attenuation curve to the intrinsic spectrum and comparing the resulting flux to the \skirt{} output. This provides a more stringent test than curve reconstruction alone: since attenuation enters exponentially, errors in $A_\lambda/A_V$ at high $A_V$ translate into disproportionately larger flux errors than equivalent residuals at low $A_V$, making flux recovery the more physically relevant metric.

To highlight trends across different attenuation regimes, we group lines of sight into eight bins of $\Av$. This is motivated by the known dependence of the shape of the attenuation curve on optical depth: low-$\Av$ sightlines tend to exhibit more pronounced UV bumps and steeper FUV slopes, while higher-$\Av$ regimes are characterized by smoother and grayer attenuation curves \citep{Salim20,Sommovigo_2025}.

The results are shown in \cref{Fig:summary_plot_EMCEE_MW}. The top panels display attenuation curve residuals, while the bottom panels show the resulting fractional flux error. Across all $\Av$ bins, the new parameterization consistently provides the most accurate recovery of both quantities. We note that, unlike the flux residuals, which are expressed in relative terms, attenuation curve residuals are shown as absolute differences in the normalized quantity $A_\lambda/\Av$. This choice avoids artificially amplifying deviations at long wavelengths, where $A_\lambda/\Av \to 0$, and more directly reflects errors in the shape of the attenuation curve, which governs how attenuation propagates into flux once $\Av$ is fixed.

The largest discrepancies across all models occur at short wavelengths ($\lambda \lesssim 2500\,\angstrom$) and around the UV bump. In the low-$\Av$ regime ($\Av \lesssim 0.3$), the two-parameter model exhibits substantial residuals in this region, with typical deviations reaching $\sim 0.5$--$1$ in $A_\lambda/\Av$, corresponding to absolute attenuation differences of order $\leq 0.3$ for the sightlines considered here. The \citet{Li08} model reduces these offsets but still shows systematic structure. In contrast, the new model keeps residuals within $\sim 0.1$--$0.2$ across the same wavelength range.

\begin{figure*}[t]
\centering

\begin{minipage}[c]{0.02\textwidth}
  \centering
  \rotatebox{90}{\textbf{MW dust}}
\end{minipage}%
\hspace{0.01\textwidth}
\begin{minipage}[c]{0.9\textwidth}
  \centering
  \includegraphics[width=\linewidth]{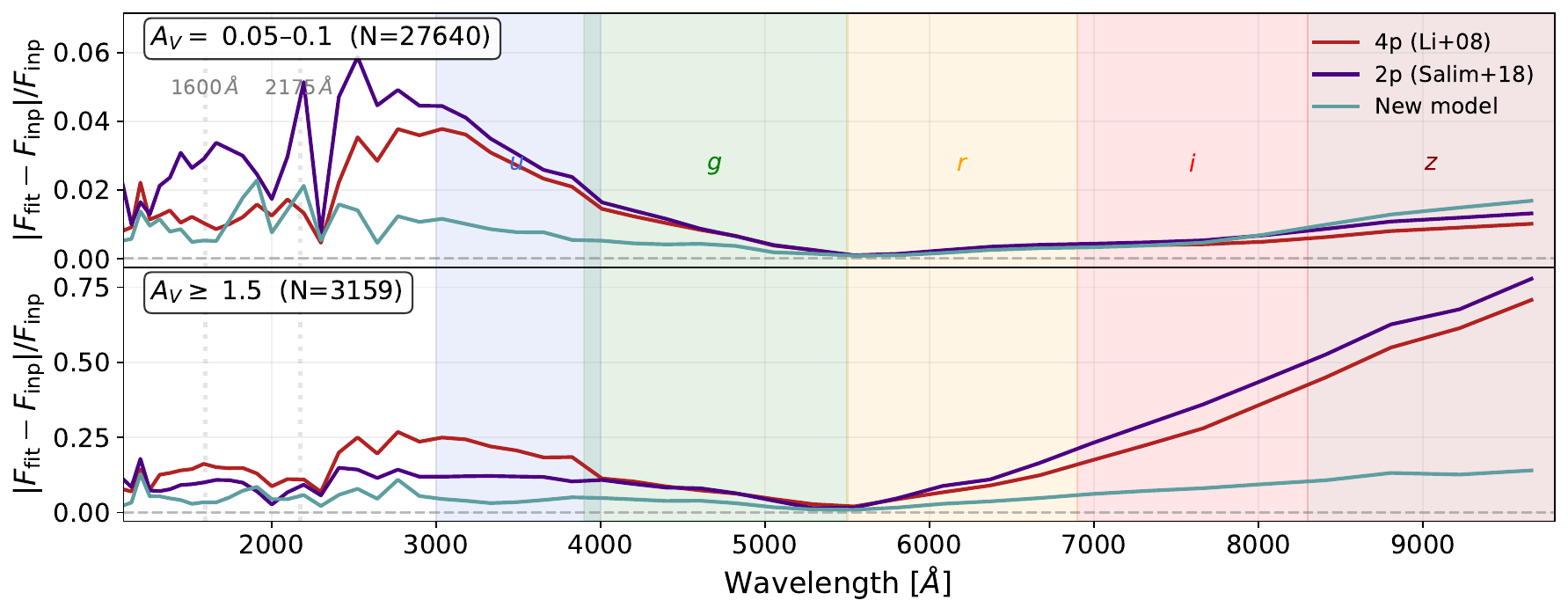}
\end{minipage}

\vspace{-2.8em}

\begin{minipage}[c]{0.02\textwidth}
  \centering
  \rotatebox{90}{\textbf{SMC dust}}
\end{minipage}%
\hspace{0.01\textwidth}
\begin{minipage}[c]{0.9\textwidth}
  \centering
  \includegraphics[width=\linewidth]{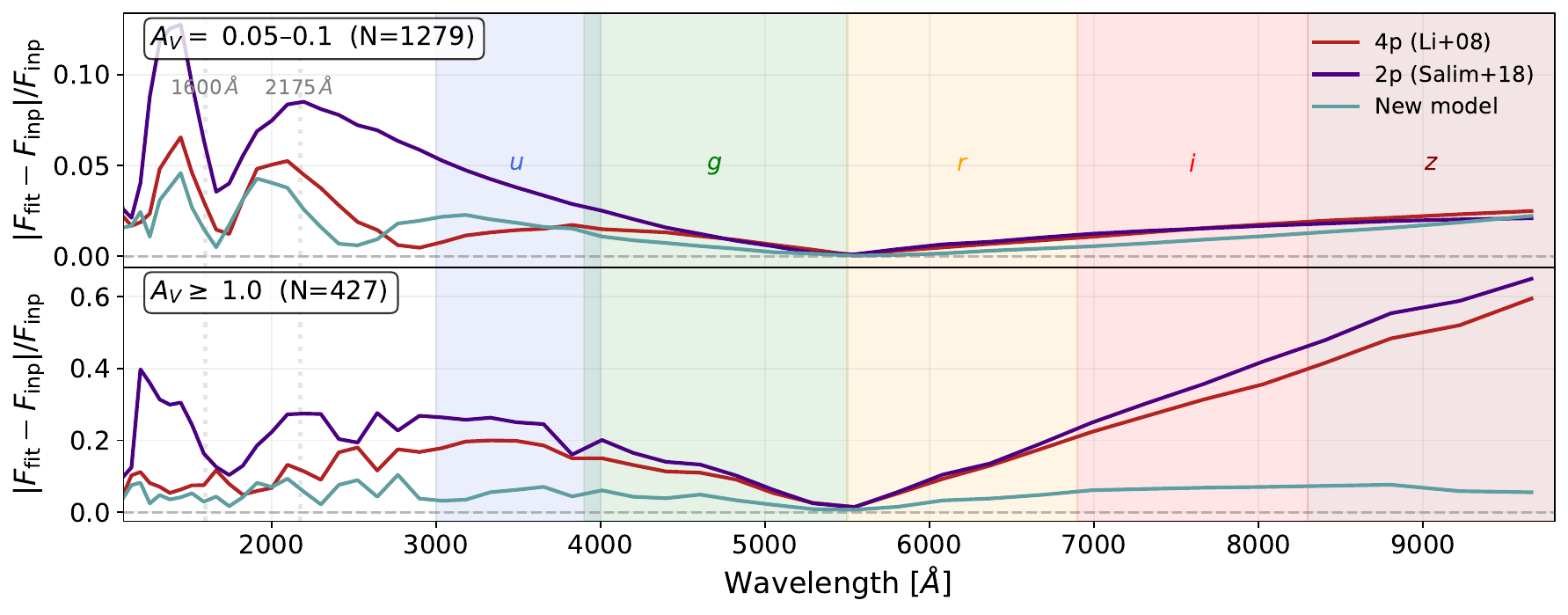}
\end{minipage}

\vspace{-2.8em}

\begin{minipage}[c]{0.02\textwidth}
  \centering
  \rotatebox{90}{\textbf{Stellar dust}}
\end{minipage}%
\hspace{0.01\textwidth}
\begin{minipage}[c]{0.9\textwidth}
  \centering
  \includegraphics[width=\linewidth]{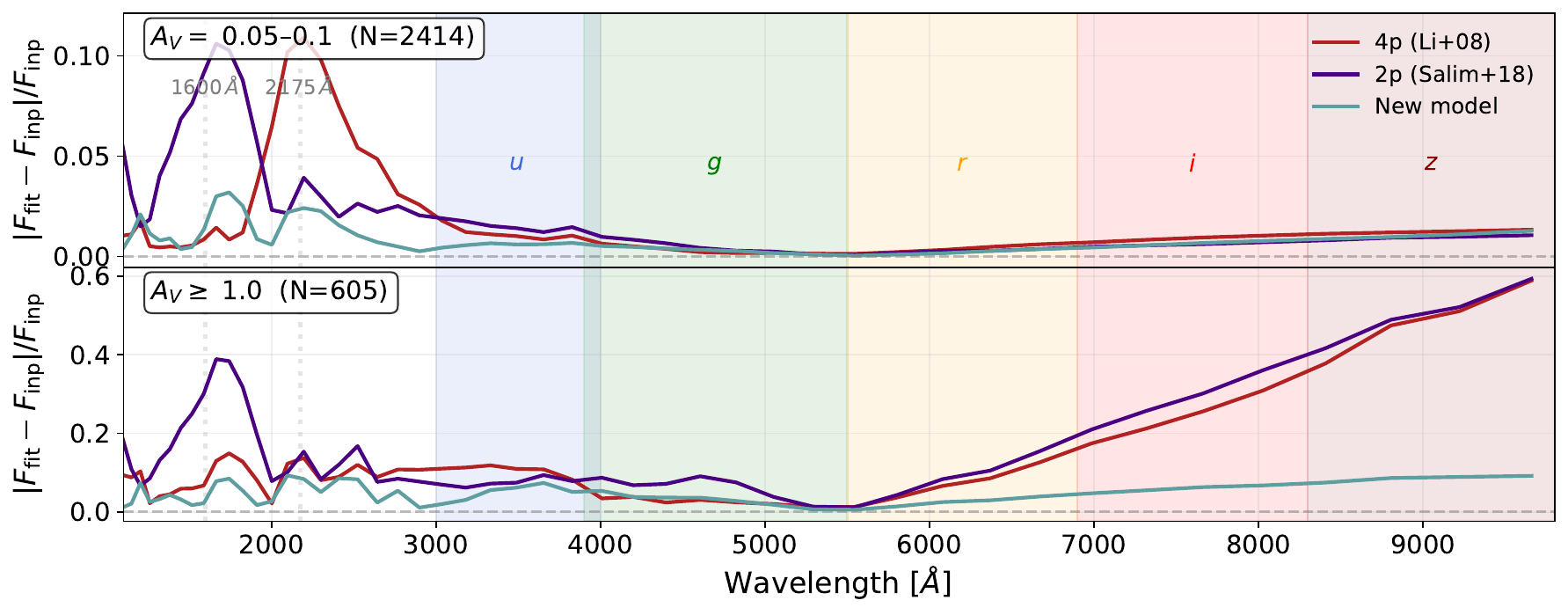}
\end{minipage}

\caption{Modulus of the fractional difference in recovered observed (attenuated) flux when relying on the best-fit attenuation curve from the three different functional forms explored here (see legend) for MW dust mixture (top row) and SMC dust (bottom row). Each panel shows the median across all lines of sight of galaxies within the lowest and highest $\Av$ bins in the simulated galaxy sample; the number of galaxies and range of $\Av$ values are reported in the top-left corner of the plots. The model proposed in this work achieves consistently smaller errors across $\Av$, wavelength and dust mixture than those from the literature.
}
\label{Fig:summary_plot_EMCEE_detail}
\end{figure*}

While these differences in normalized attenuation are large at low $\Av$, their impact on flux recovery remains limited, as the overall attenuation is small. To better isolate wavelength-dependent trends and facilitate comparison across different dust mixtures, we also show in \cref{Fig:summary_plot_EMCEE_detail} the median absolute fractional flux residuals for representative low- and high-$\Av$ regimes. This is confirmed in the top row of \cref{Fig:summary_plot_EMCEE_detail}, where all models achieve fractional flux errors $\lesssim 5\%$ at $\Av \sim 0.05$--$0.1$, despite significant differences in attenuation curve space at short UV wavelengths.

The situation reverses at higher attenuation. For $\Av \gtrsim 1$, discrepancies in the optical ($\lambda \gtrsim 5000\,\angstrom$) dominate the flux error budget. In this regime, both the \citet{Li08} and \citet{Salim18} parameterizations systematically overpredict the attenuation at long wavelengths, leading to flux errors that increase steadily toward the red, reaching $\sim 40$--$80\%$ at $\lambda \sim 9000\,\angstrom$ (\cref{Fig:summary_plot_EMCEE_detail}, top row). These trends reflect the inability of these models to reproduce the UV shape and the optical slope simultaneously.

The new parameterization substantially mitigates these biases. Across all $\Av$ bins, it maintains flux errors at the $\sim 5$--$15\%$ level over the full wavelength range, with no strong systematic drift toward either the UV or optical. This improvement is particularly evident at high $\Av$, where the new model recovers significantly flatter attenuation curves and avoids the systematic steepening seen in the other parameterizations.

In summary, these results show that the improved performance of the new model is driven by its ability to simultaneously capture the UV bump region and the optical slope, which are the primary sources of error in existing parameterizations\footnote{
We have verified that these conclusions are unchanged when performing the fits directly in flux space, i.e.\ by fitting the observed flux using the intrinsic and dust-attenuated \skirt{} output spectra, 
 rather than fitting the derived attenuation curves. In this case, $\Av$ is treated as an additional free parameter in all parameterizations. This setup more closely resembles standard SED fitting approaches and yields consistent relative performance across models.
}. In the next subsection, we test whether these conclusions hold when varying the underlying dust composition.

\begin{figure*}[t]
\centering

\begin{minipage}[c]{0.02\textwidth}
  \centering
  \rotatebox{90}{\textbf{SMC dust}}
\end{minipage}%
\hspace{0.01\textwidth}
\begin{minipage}[c]{0.95\textwidth}
  \centering
  \includegraphics[width=\linewidth]{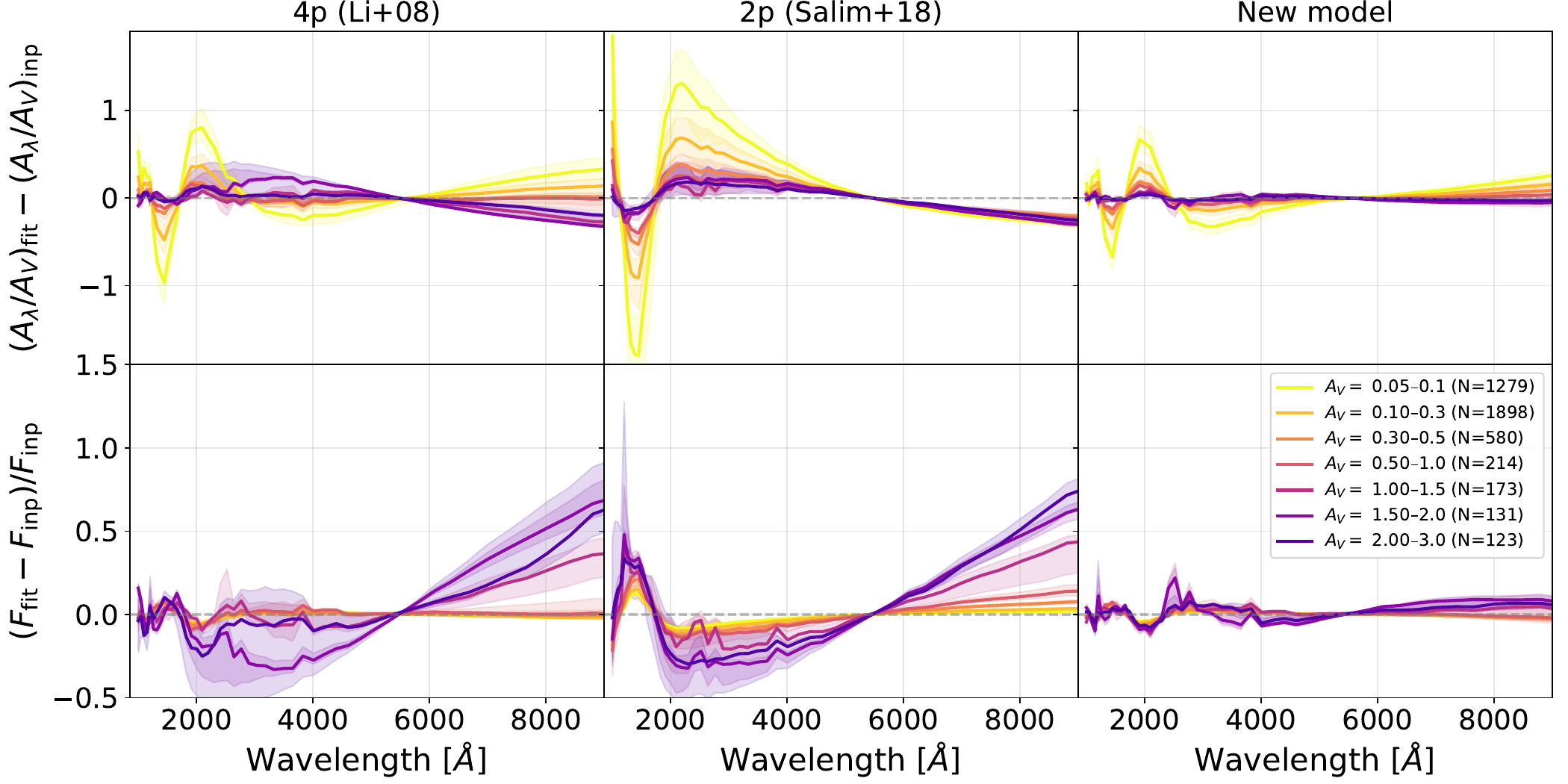}
\end{minipage}

\vspace{-2.65em}

\begin{minipage}[c]{0.02\textwidth}
  \centering
  \rotatebox{90}{\textbf{Stellar dust}}
\end{minipage}%
\hspace{0.01\textwidth}
\begin{minipage}[c]{0.95\textwidth}
  \centering
  \includegraphics[width=\linewidth]{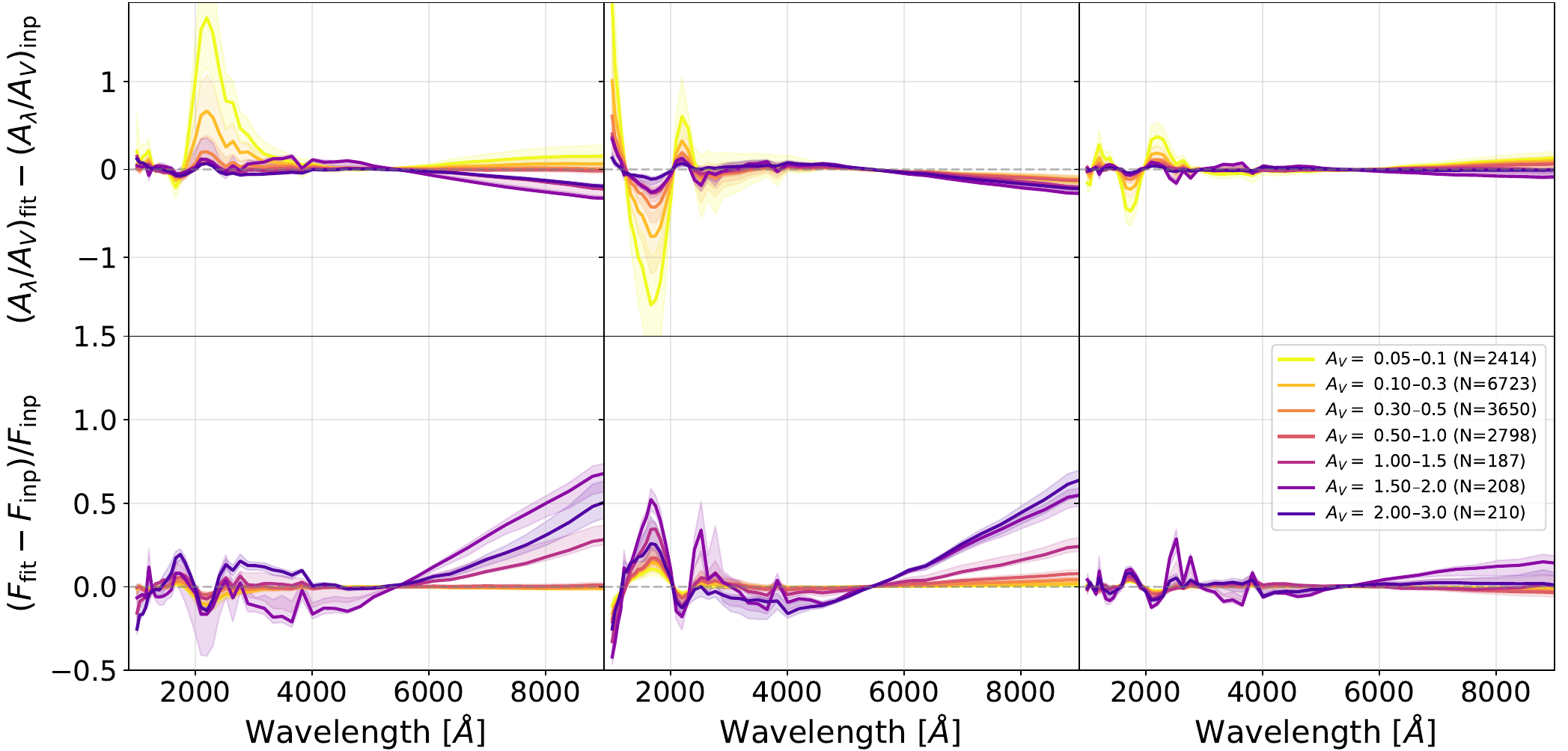}
\end{minipage}
    \caption{Same as \cref{Fig:summary_plot_EMCEE_MW}, but for the SMC (top) and stellar (bottom) dust mixtures. 
    Using our model yields an improved fit, even for dust models it was not trained on.
    }
\label{Fig:summary_plot_EMCEE_SMC}
\end{figure*}

\subsection{Impact of dust composition on model performance and attenuation curve parameters distribution}

To test the robustness of the different functional forms to the assumed dust mixture, we repeat the radiative transfer calculations for a subsample of galaxies adopting two alternative dust compositions: an SMC-like model \citep{Weingartner01} and a stellar dust \citep{Hirashita19} model (see Section~\ref{sec:data} for more details). As described in Section~\ref{sec:data}, we rerun the \skirt{} simulations for 482 galaxies selected to uniformly sample the $\log M_\star$–$\Sigma_{\rm SFR}$ plane using these dust mixtures.

The SMC and stellar dust models probe distinct regions of attenuation curve space. The SMC mixture  -- in the \citep{Draine03} model -- lacks PAHs, and therefore does not exhibit a prominent $2175\,\angstrom$ bump, while still retaining a large fraction of small grains that efficiently absorb at short wavelengths, resulting in steeper UV slopes. 
In contrast, the stellar dust model is significantly depleted in small grains, leading to a much smaller fraction of particles with sizes $a \lesssim 0.17\,\mu$m ($\sim 30\%$ by mass, compared to $\sim 60\%$ and $\sim 70\%$ for MW and SMC dust, respectively). Since grains with sizes comparable to or smaller than the wavelength are the most efficient absorbers in the UV, this depletion suppresses the wavelength dependence of the attenuation, producing systematically flatter (grayer) curves. 
Both stellar and SMC-like attenuation curves occupy lower-dimensional regions of attenuation curve space compared to the Milky Way case due to the lack of strong features, but in qualitatively different directions: the SMC model enhances UV steepness, while the stellar dust model flattens the overall attenuation law.

We now assess whether the relative performance of the different parameterizations is preserved when moving away from the Milky Way regime. This provides a stringent test of whether the improved performance of the new model reflects a genuine ability to capture attenuation curve structure for a variety of dust models, rather than a tuning to MW-like features.
The results are shown in \cref{Fig:summary_plot_EMCEE_SMC}, which adopts the same visualization as \cref{Fig:summary_plot_EMCEE_MW} to enable a direct comparison across dust mixtures. Across all $\Av$ bins, the new parameterization continues to provide the most accurate recovery of the attenuation curves.

In attenuation curve space (top panels), the nature of the discrepancies changes compared to the MW case.
For the SMC dust model, the absence of a UV bump removes one of the main sources of localized residual structure. 
However, the significant improvement provided by the new model persists at short wavelengths,
where the two-parameter model shows excursions approaching the limits of the plotted range ($\sim 0.5$ in magnitude) in $A_\lambda/\Av$. More importantly, systematic offsets in the overall slope remain at longer wavelengths, indicating that the dominant source of error is not tied to the presence of the $2175\,\angstrom$ bump. In contrast, the new parameterization maintains residuals at the $\sim 0.1$--$0.2$ level across the full wavelength range.
This is the same as for the MW dust mixture (which was used to derive the functional form), demonstrating good extrapolation behavior to new dust mixtures.

The stellar dust case provides a more stringent test of model flexibility. In this regime, the two-parameter model fails to reproduce the flattening at short wavelengths, leading to large residuals (approaching $\sim \pm 1$ in magnitude) at low $\Av$. This reflects the inability of a single slope parameter to simultaneously describe both the optical regime and the suppressed UV attenuation. The \citet{Li08} model performs better in capturing the overall slope, but still exhibits systematic residuals at short wavelengths due to the coupling between parameters governing the UV slope and the bump region, which can introduce spurious curvature even when no bump is present. As a result, neither of the literature parameterizations can simultaneously reproduce the global slope and the UV flattening in the stellar dust case.

To quantitatively assess how these differences propagate into observable quantities, we now examine the reconstructed flux in representative low- and high-$\Av$ regimes.
As shown in \cref{Fig:summary_plot_EMCEE_detail}, systematic trends in flux space are qualitatively consistent across MW, SMC, and stellar dust. At long wavelengths ($\lambda \gtrsim 5000$\,\angstrom), both the \citet{Li08} and \citet{Salim18} parameterizations exhibit systematic errors that increase with $\Av$, reaching $\sim 30$--$70\%$ at $\lambda \sim 9000$\,\angstrom\ for $\Av \gtrsim 1$. This reflects a persistent mismatch in the global optical slope, which dominates the flux error budget.

At shorter wavelengths, the dominant source of error depends on the parameterization and dust mixture. The two-parameter model shows significant flux discrepancies in the FUV and UV ($\sim 10$--$20\%$), reflecting its inability to capture curvature and local structure, even in the absence of a bump. The four-parameter \citet{Li08} model reduces these discrepancies, but still exhibits systematic errors due to parameter coupling, which allows spurious bump-like features to emerge and limits its flexibility in the stellar dust (or in general flat attenuation curve) regimes.
In contrast, the new parameterization maintains flux errors at the $\sim 5$--$15\%$ level across the full wavelength range, with no strong systematic drift. 

Taken together, these results show that while changes in dust composition modify the detailed shape of attenuation curves, they do not remove the dominant sources of error in existing parameterizations. The improved performance of the new model instead stems from its ability to simultaneously capture both the optical slope and the UV behavior across the full range of attenuation curve shapes.

\begin{figure*}
    \centering
    \includegraphics[width=0.93\linewidth]{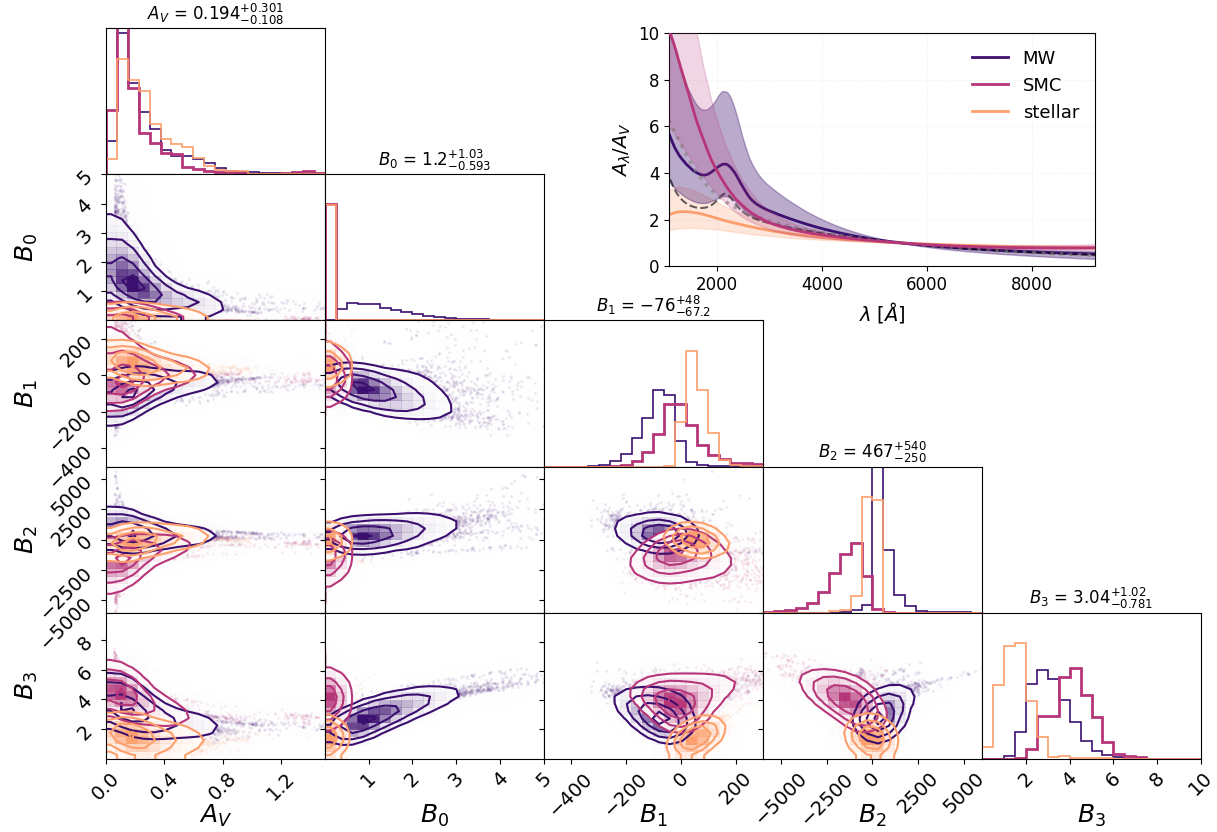}    
    \caption{Corner plot of the best-fit parameters of our attenuation curve model (\cref{eq:final_model}) for the 487 galaxies from TNG50 and TNG100, each sampled along 51 lines of sight and fitted assuming three dust mixtures (Milky Way, SMC, and stellar dust). Colors indicate different dust compositions. The inset shows the corresponding median attenuation curves, with shaded regions indicating the 16th--84th percentile range. For reference, the observed Milky Way \citep{Cardelli89} and SMC \citep{Prevot84} extinction curves are also shown (black dashed and gray dotted lines, respectively). This comparison illustrates the impact of dust composition on both the best-fit parameter distributions and the resulting attenuation curve shapes.}
    \label{fig:corner_allpar_all_dust}
\end{figure*}

In addition to differences in model performance, it is important to assess how the inferred parameters themselves respond to changes in dust composition. This is shown in \cref{fig:corner_allpar_all_dust}, where we compare the distributions of best-fit parameters for the same set of 487 galaxies across the MW, SMC, and stellar dust models.

We find that the parameters shift systematically across dust mixtures, even for an identical underlying galaxy population. More generally, the corner plot shows that dust mixtures are not only separated in the one-dimensional parameter distributions, but also occupy distinct regions in several two-dimensional parameter planes. This is particularly evident in several two-dimensional parameter planes, most clearly in $B_0$--$B_1$, $B_1$--$B_3$, $B_0$--$B_3$, and $B_2$--$B_3$, where different dust mixtures occupy distinct regions of parameter space. These separations suggest that combinations of parameters, rather than individual parameters alone, may be preferentially sensitive to changes in dust composition. This could provide a way to disentangle variations in grain properties from line-of-sight and radiative-transfer effects, although this interpretation remains tentative given the limited number of dust mixtures explored here.

This naturally raises the question of whether these apparent separations reflect genuine, independent sensitivity to dust composition, or instead arise from internal degeneracies within the parameterization. In particular, one may ask whether a reduced set of parameters would be sufficient to capture these trends, or whether the current level of complexity is required to preserve the quality of the fits across different dust regimes. We address this by quantifying the internal parameter correlations across all dust mixtures in the following section.

\subsection{Parameter degeneracies and impact of dust composition}
\begin{figure*}[t]

\includegraphics[width=1.\linewidth]{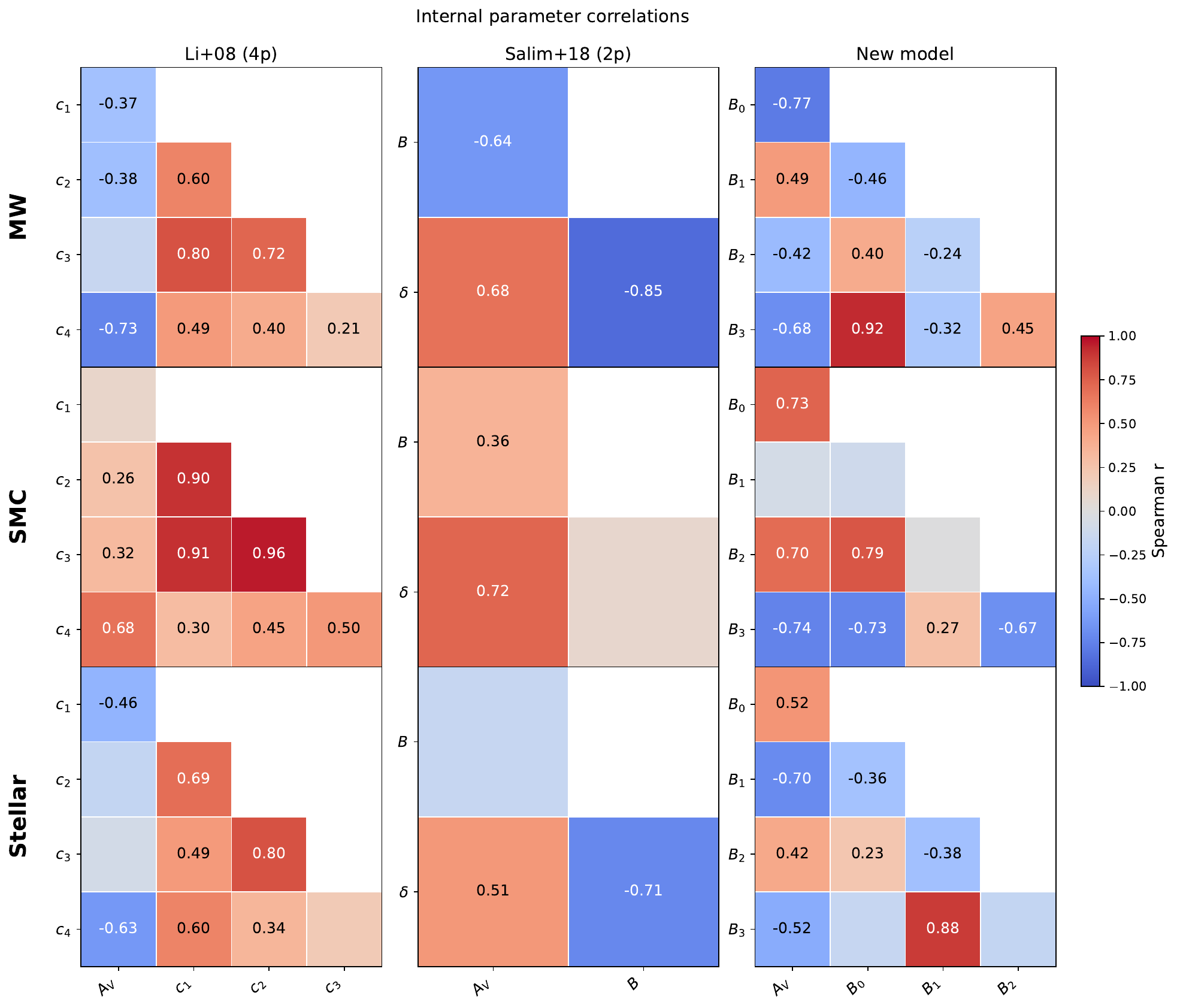}
\caption{
Internal parameter correlations (Spearman rank coefficients, $r$) for the different attenuation curve parameterizations under varying dust compositions. Each panel shows the correlations among free parameters for all three functional forms (Li+08, Salim+18, and the new model). For clarity, only the lower triangular portion of each symmetric correlation matrix is displayed. Similarly, we only print numerical values if $|r|\geq 0.2$.
The structure of the correlations differs across dust models, particularly for parameters associated with the UV bump ($c_4$, $B$, $B_0$ and $B_2$), indicating that no universal reparameterization can be defined across dust compositions.
}
\label{fig:internal_corr_dust}
\end{figure*}

A natural question is whether the improved performance of the new parameterization simply reflects an increased flexibility and whether some of its parameters are effectively redundant and could be removed without loss of accuracy. To address this, we examine the internal parameter correlations across all parameterizations and dust mixtures, shown in \cref{fig:internal_corr_dust}.

All three parameterizations exhibit significant internal correlations, indicating that none of them is fully orthogonal. Importantly, our new model is not exempt from this behavior. Quantitatively, when considering correlations among the shape parameters only, we find that $\sim 39\%$ of parameter pairs in our model have Spearman correlation coefficient $|r| < 0.3$, compared to $\sim 11\%$ for \citet{Li08} and $\sim 33\%$ for \citet{Salim18} (corresponding to $7/18$, $2/18$, and $1/3$ pairs, respectively). This indicates a somewhat more orthogonal representation, although not a fully decoupled one.

However, two key features emerge. First, the strength and structure of the correlations vary substantially with dust composition. In particular, correlations generally become stronger in the SMC and stellar cases, reflecting the lower intrinsic dimensionality of attenuation curve space in the absence of prominent spectral features. Second, these correlations do not follow a consistent pattern across dust models: parameter couplings that appear strong in one dust regime weaken or change sign in another.

This has direct implications for model complexity. The dust mixture dependent nature of the correlations implies that no subset of parameters can be safely eliminated in a way that remains valid across different dust compositions. In other words, even if attenuation curves occupy a lower-dimensional space for a given dust model, this reduction does not translate into a universal simplification of the parameterization. Importantly, we have only explored three dust mixtures here; the true diversity of grain populations in galaxies across cosmic time is likely far broader, making the need for a flexible, general parameterization even more pressing.

In conclusion, these results show that the improved performance of the new model is not driven by overfitting, but by a more flexible and better-conditioned parameterization. Retaining all parameters is necessary to capture the diversity of attenuation curve shapes across dust mixtures, and any reduced parameterization would inevitably fail to generalize across regimes. In this context, the strength of the new functional form lies in its ability to adapt to both higher- and lower-dimensional regimes without requiring prior knowledge of the underlying dust composition.

\section{Correlations with galaxy properties: robustness to parameterization and dust composition}\label{sec:galaxy_correl}

In \citet{Sommovigo_2025}, we investigated the connection between attenuation curve shape and galaxy properties using the \citet{Li08} parameterization applied to synthetic attenuation curves from the TNG50 simulations. 
The galaxy properties considered include star-formation rates averaged over $10\,{\rm Myr}$ and $100\,{\rm Myr}$ (${\rm SFR}_{10}$ and ${\rm SFR}_{100}$), their corresponding specific star-formation rates (${\rm sSFR}_{10}$ and ${\rm sSFR}_{100}$), the logarithms of the mass-weighted stellar age, stellar, gas, and dust masses ($
\log_{10}(M_\ast/M_{\odot})$, $\log_{10}(M_{\rm g} /M_{\odot})$, $ \log_{10}(M_{\rm dust} /M_{\odot})$), gas metallicity ($Z_{\rm g}$), and a set of structural quantities describing the spatial distribution of stars and gas. 
These include the characteristic radii of young and old stellar populations ($r_{\ast,{\rm y}}$ and $r_{\ast,{\rm o}}$), as well as the half-mass radii of all gas and star-forming gas ($r_{\rm g}$ and $r_{\rm g,SF}$). We additionally consider ratios of these quantities, surface densities (e.g.\ $\Sigma_{\rm SFR}$, $\Sigma_{\rm g}$, $\Sigma_{\rm dust}$), and deviations from the Kennicutt--Schmidt relation parameterized through $k_{\rm s}$ quantities \citep{Kennicutt98}.  A complete list of the quantities considered, together with their definitions, is provided in Table~4 of \citet{Sommovigo_2025}. 

Here, we extend this analysis along two complementary directions. First, we test whether the inferred correlations depend on the specific functional form adopted to describe the attenuation curve. Second, we assess their robustness to changes in the underlying dust composition by repeating the analysis for Milky Way, SMC-like, and stellar dust mixtures.

Compared to \citet{Sommovigo_2025}, we additionally include the line-of-sight inclination relative to the galaxy disk plane among the galaxy properties considered, $\theta_{\rm inc}$, defined as the angle between the line of sight and the angular momentum vector of the stellar component, computed within twice the stellar half-mass radius. 
We also explicitly explore the case in which attenuation curves from different dust mixtures are combined into a single sample.
This allows us to distinguish correlations that remain robust when the dust composition is unknown from those that depend significantly on the microscopic grain properties.

Throughout this section, we compute correlations using all individual lines of sight independently, following the same approach adopted throughout this work. As already discussed in \citet{Sommovigo_2025}, averaging attenuation curves over lines of sight strengthens the correlations with galaxy properties by suppressing line-of-sight-driven variations. The line-of-sight analysis adopted here therefore provides a more conservative estimate of the predictive power of galaxy properties in realistic observational settings.

We begin by comparing the correlations obtained across different attenuation curve parameterizations, and then examine how these trends change when varying the dust mixture.
When comparing parameterizations, since we find that correlations are stronger with a single dust mixture, we only report the results for a single dust mixture rather than for the combined sample containing all dust mixtures. For this, we consider the MW mixture but have verified that our conclusions are the same for the stellar and SMC dust mixtures, and for a combined sample of all dust mixtures.

\begin{figure}[t]
\centering
\includegraphics[width=1.\linewidth]{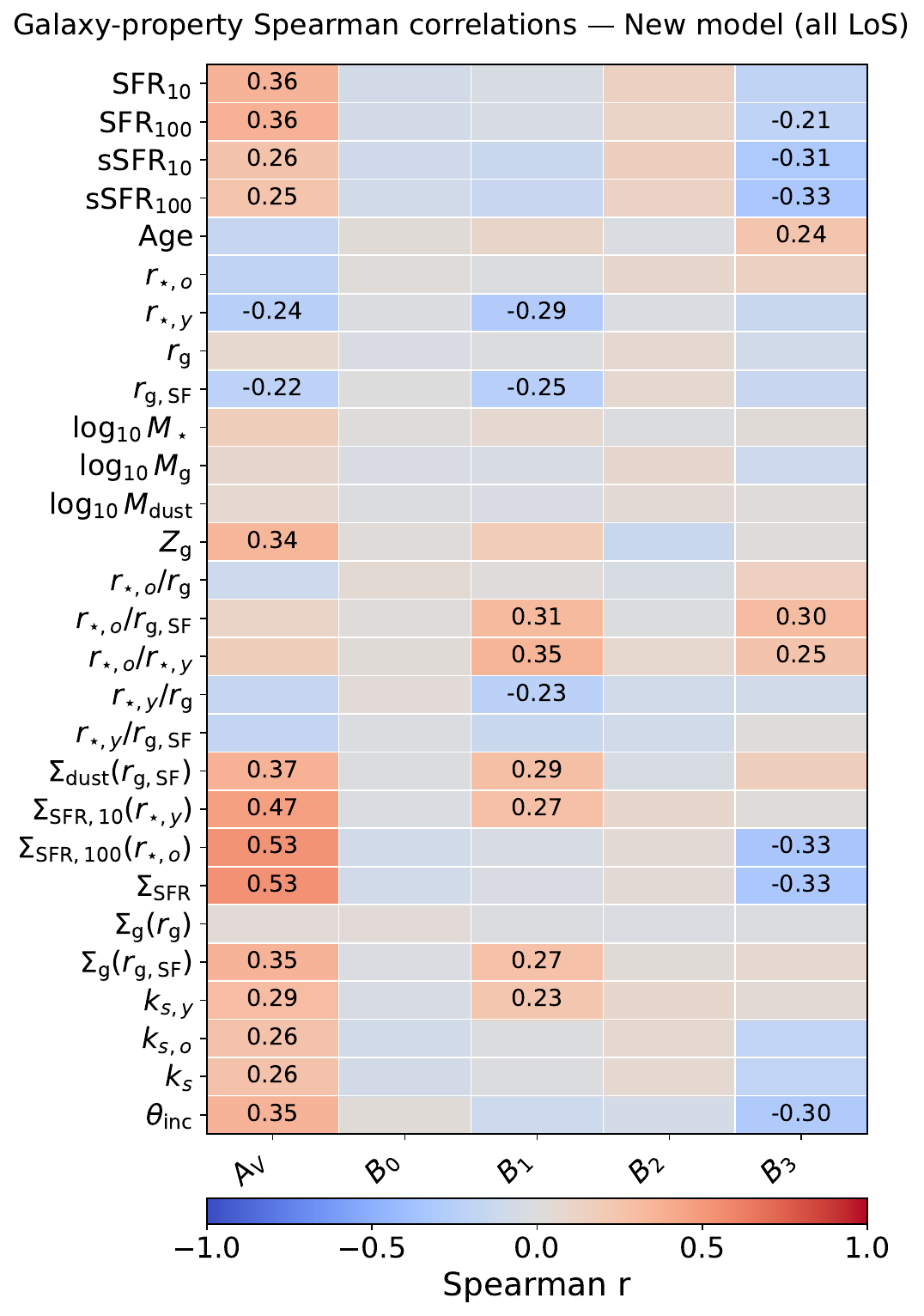}
\caption{
Correlation matrices between attenuation curve parameters and galaxy properties for our new parameterization, computed using all lines of sight. 
Here we show the case where attenuation curves from different dust mixtures are combined and treated as additional lines of sight, and in Fig.~\ref{fig:correlations_dust_mixes} we separate the dust mixtures. 
For clarity, we only print Spearman correlation coefficients, $r$, that satisfy $|r| \geq 0.2$.
The overall structure of the correlations is preserved across dust mixtures, with $\Av$ and $\Sigma_{\rm SFR}$ remaining the dominant drivers. Combining dust mixtures introduces additional scatter that erases correlations involving bump-related parameters. 
}
\label{fig:correlations_newmodel_dust_combined}
\end{figure}

\subsection{Robustness to parameterization}

In the Appendix (the left panel of \cref{fig:correlations_dust_mixes}) we summarize the Spearman correlation coefficients between the attenuation curve parameters of our new functional form and galaxy properties for the MW dust mixture, considering all individual lines of sight independently. 
The corresponding results for the \citet{Li08} and \citet{Salim18} parameterizations are also presented in the Appendix (\cref{fig:correlations_other_models_MW}) and lead to qualitatively consistent conclusions.

\begin{figure}[t]
\centering
\includegraphics[width=\linewidth]{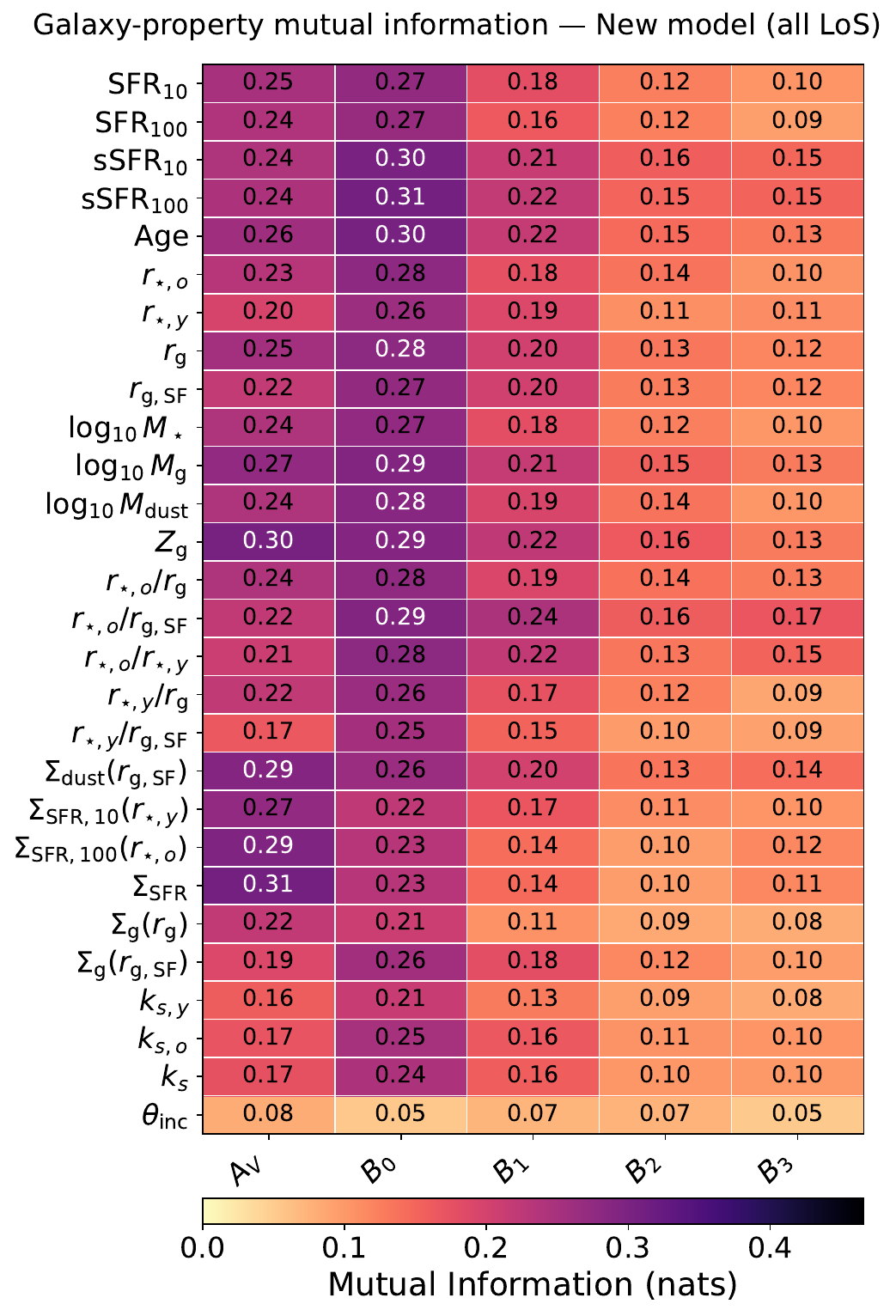}
\caption{
\cref{fig:correlations_newmodel_dust_combined}, but using \textbf{mutual information} instead of Spearman rank coefficients. Here we show the case where attenuation curves from different dust mixtures are combined and treated as additional lines of sight, and in Fig.~\ref{fig:correlations_MI_dust_mixes} we separate the dust mixtures.  Yellow regions correspond to nearly independent quantities, while darker/purple shades indicate stronger statistical dependence. Unlike Spearman coefficients, which only capture monotonic trends, mutual information is sensitive to more general nonlinear and non-monotonic relationships between attenuation curve parameters and galaxy properties. 
}
\label{fig:correlations_MI}
\end{figure}

For a fixed dust mixture, the overall structure of the correlations is robust to the choice of functional form. Parameters controlling analogous attenuation curve features show qualitatively similar correlation patterns across parameterizations, once differences in sign convention are accounted for. For example, the \citet{Salim18} slope parameter $\delta$ and our $B_3$ both encode large-scale changes in the UV-to-optical slope, but with opposite sign conventions: increasing $\delta$ produces flatter attenuation curves, whereas increasing $B_3$ steepens the UV rise. Their correlations with galaxy properties therefore appear with opposite signs but trace the same underlying dependence\footnote{We do not interpret the \citet{Li08} parameter $c_2$ as a one-to-one slope analogue, since its effect on the attenuation curve is non-monotonic and partly degenerate with the other attenuation curve parameters.}.

Parameters regulating the broad UV shape and FUV rise ($c_1$ and $c_3$ in \citealt{Li08}, $\delta$ in \citealt{Salim18}, and $B_1$ and $B_3$ in our parameterization) also show qualitatively consistent trends, correlating most strongly with $\Sigma_{\rm SFR}$, sSFR, and inclination, typically at the level of $|r| \sim 0.3$--$0.5$. 

Similarly, the parameters primarily associated with the UV bump amplitude ($c_4$, $B$, and $B_0$) all show negative correlations with $\Sigma_{\rm SFR}$, sSFR, and inclination for the MW dust mixture. It is worth noting, however, that these trends are likely partly secondary, as the bump parameters are themselves strongly anti-correlated with $\Av$ ($r \sim -0.7$ across all parameterizations), while $\Av$ positively correlates with the same galaxy properties.

Taken together, these results indicate that the dominant relationships linking attenuation curve shape to galaxy properties are largely physical rather than artifacts of a specific parameterization.

It is worth noting that, despite residual internal degeneracies, our parameterization exhibits a larger fraction of non-negligible correlations with galaxy properties than the \citet{Li08} model while retaining the flexibility of a four-parameter description. Considering entries with $|r| \gtrsim 0.2$, only $\sim 30$--$40\%$ of the \citet{Li08} correlations are non-negligible, compared to $\sim 45$--$55\%$ for our model. At the same time, our parameterization reaches a level comparable to the two-parameter \citet{Salim18} model, for which $\sim 40$--$60\%$ of entries are non-negligible. This indicates that our functional form combines the descriptive power of higher-dimensional parameterizations with a cleaner and more physically interpretable mapping onto galaxy properties.

\subsection{Robustness to dust composition}
We now repeat the same analysis considering different dust mixtures to assess the robustness of the attenuation parameter--galaxy property correlations to changes in dust composition focusing only on our parameterization. As discussed above, this functional form provides the best overall reconstruction accuracy across all dust compositions while also exhibiting a cleaner mapping onto galaxy properties than the \citet{Li08} parameterization, despite retaining a four-parameter description. This makes it the most suitable framework for assessing which attenuation--galaxy property correlations are genuinely robust to changes in dust composition.
The resulting correlation matrices are shown in \cref{fig:correlations_newmodel_dust_combined,fig:correlations_dust_mixes}, where the different panels correspond to MW, SMC-like, stellar, and combined dust mixtures.

Across all dust mixtures, $\Av$ remains the attenuation parameter most strongly correlated with galaxy properties. The strongest trends are found with star-formation rate surface densities — in particular $\Sigma_{\rm SFR,10} (r_{\star,y})$, which reaches $r \sim 0.5$--$0.6$ — followed by dust surface density within the star-forming gas, gas-phase metallicity, and line-of-sight inclination (typically $r \sim 0.3$--$0.4$). That $\Sigma_{\rm SFR}$ normalized by the young stellar half-mass radius correlates most strongly with $\Av$ is physically intuitive: it jointly encodes the intensity of the UV radiation field and the spatial concentration of its sources relative to the dust distribution, both of which govern the effective optical depth along any given line of sight.

At fixed dust mixture, internal correlations between $\Av$ and the free attenuation curve parameters are often stronger than the direct correlations between the shape parameters and galaxy properties, as already pointed out in \citet{Sommovigo_2025} for the MW dust mixture alone. However, several of these internal correlations weaken substantially or disappear once attenuation curves from different dust mixtures are analyzed jointly (see \cref{fig:correlations_newmodel_dust_combined}). This demonstrates that the covariance structure of attenuation curve parameters is itself dust mixture dependent, consistent with the changing geometry and dimensionality of attenuation curve parameter space discussed in the previous sections.

Within our parameterization, $B_1$ and $B_3$ exhibit the most robust correlations with galaxy properties across dust mixtures, particularly with $\Sigma_{\rm SFR}$, sSFR, and inclination, reaching $|r| \sim 0.4$--$0.5$. These trends remain visible (albeit weakened) even when combining all dust mixtures, indicating that they trace aspects of attenuation curve shape that are comparatively insensitive to the microscopic dust composition. Physically, both parameters regulate the large-scale UV-to-optical behavior of the attenuation curve, with $B_3$ controlling large scale slope and $B_1$ modulating the UV one. Among the shape parameters, $B_1$ shows the strongest individual Spearman correlation in the combined dust mixture sample, driven primarily by structural ratios that quantify the relative spatial extent of young and old stellar populations ($r_{\star,\mathrm{o}}/r_{\star,\mathrm{y}}$, $r_{\star,\mathrm{o}}/r_{\mathrm{g,SF}}$). This is physically expected \citep[see also][]{Narayanan18,Matsumoto26}: $B_1$ modulates the FUV slope of the attenuation curve, which is
most sensitive to the spatial offset between UV-emitting young stars and the surrounding dust. When young stars are more centrally concentrated than the old population, they are embedded in higher-column-density regions, steepening the FUV attenuation relative to the optical. Because this mechanism is primarily geometric, the correlation persists across dust mixtures, unlike the bump-sensitive parameters $B_0$ and $B_2$ whose correlations are more dust-composition dependent.

In contrast, $B_2$, which regulates the curvature around the UV bump transition region, shows only weak correlations with galaxy properties ($|r| \lesssim 0.2$--$0.3$) that -- unsurprisingly -- disappear almost entirely once different dust mixtures are combined. This suggests that $B_2$ primarily captures variations associated with line-of-sight geometry and dust mixture specific structure in attenuation curve space, rather than global galaxy properties.
The behavior of $B_0$ is qualitatively different. As expected, it is only significantly non-zero for the MW dust mixture, where it shows negative correlations with $\Sigma_{\rm SFR}$, sSFR, and inclination. However, these trends are likely at least partially secondary, since $B_0$ is itself strongly anti-correlated with $\Av$, implying that highly attenuated lines of sight systematically exhibit weaker UV bumps.

To assess whether some of the weakened correlations in the combined dust mixture sample arise from the loss of monotonicity rather than from a complete disappearance of the underlying dependence, we repeated the same analysis using mutual information instead of Spearman rank coefficients (\cref{fig:correlations_MI,fig:correlations_MI_dust_mixes}). Mutual information measures the reduction in uncertainty (entropy) on one quantity provided by knowledge of another, and is therefore sensitive to general nonlinear and non-monotonic relationships that rank-based statistics miss. We note that mutual-information values do not have a fixed interpretive scale: they depend on the marginal entropy of the variables involved, so they should be compared across panels and parameters within the figure rather than interpreted in absolute terms.            
     
The mutual-information analysis broadly confirms the dominant Spearman trends — metallicity, $\Sigma_{\rm SFR}$, and sSFR remain the strongest predictors of $\Av$, $B_1$, and $B_3$ — while also revealing      
additional structure. Most notably, $B_0$ and $B_2$ retain substantial mutual information with metallicity, stellar mass, gas mass, and characteristic galaxy sizes, even in regimes where their Spearman correlations become weak or vanish. This is especially striking for the bump-strength parameter $B_0$: its Spearman correlations are significant only for the MW dust mixture, where a UV bump is present, yet the mutual-information analysis reveals residual dependence on galaxy properties across all dust mixtures, implying that galaxy properties encode information about UV-structure variations in a non-monotonic, dust mixture dependent way. More broadly, global structural quantities such as stellar mass, gas mass, and galaxy sizes gain importance in the mutual-information analysis relative to the Spearman one, suggesting that they shape attenuation curve properties through relations that depend jointly on dust composition and viewing geometry.

Taken together, these analyses indicate that while the detailed mapping between attenuation curve parameters and galaxy properties depends on the assumed dust composition, the dominant physical dependencies — star-formation rate, metallicity, stellar and SF gas morphology, and viewing geometry — remain broadly preserved across dust mixtures. The mutual-information results further show that part of this information persists even when monotonic trends weaken, becoming increasingly nonlinear for UV-sensitive attenuation features. Overall, macroscopic galaxy properties and line-of-sight-dependent quantities primarily regulate the optical-to-NUV attenuation behavior, while microscopic dust composition more strongly shapes the detailed FUV rise and UV bump structure.

\subsection{Predicting attenuation curves from galaxy properties via symbolic regression}\label{sec:sr_galaxy_prop}

Using the optimized parameters from Section~\ref{sec:symreg}, we now use symbolic regression with \operon{} to derive analytic expressions that predict attenuation curve parameters from global galaxy properties. 
We do this for two reasons: firstly, we use these expressions to gain further insights into the correlations between parameters and galaxy properties, but also to provide a method for constructing realistic attenuation curves for simulated galaxy catalogs which contain such correlations.
For $\Av$, we search for functions depending only on galaxy properties. For the attenuation curve parameters $B_1$ and $B_3$, we instead search for functions of both $\Av$ and the galaxy properties, since part of the observed correlation with galaxy properties is expected to arise indirectly through correlations with $\Av$.
In \cref{fig:correlations_newmodel_dust_combined}, we saw that $B_0$ and $B_2$ did not strongly correlate with galaxy properties, and thus we seek functions of only the other attenuation curve parameters. 
If one allowed both $B_0$ to be a function of $B_2$ and $B_2$ to be a function of $B_0$, then predicting the pair $(B_0, B_2)$ would be challenging for generating synthetic catalogs in a hierarchical model. Empirically, we find little change when predicting $B_0$ if we do not allow its value to depend on $B_2$, hence we search for $B_0$ as a function of $B_1$, $B_3$ and $\Av$, and for $B_2$ as a function of $B_0$, $B_1$, $B_3$ and $\Av$.

All \operon{} runs used a basis set consisting of arithmetic operations ($+$, $-$, $\times$, $\div$) as well as the power operator and logarithm. We allowed a maximum expression length of 20, and a time limit of 2 hours per run. The optimization jointly considered the coefficient of determination ($R^2$) and model complexity, as quantified by the expression length. For $\Av$ and $B_0$, we model the logarithm of the parameter rather than the parameter itself, since these quantities are strictly non-negative and their distributions within the training and validation sets are more closely approximated by log-normal rather than normal distributions.
For all models we used 2000 training points with 1000 reserved for validation, and ensured that the same galaxy did not appear in both training and validation.
For runs where we searched for a single function over all dust mixtures, both training and validation sets contained approximately equal ratios of each of the three dust mixtures considered in this work.
As before, model selection is determined by comparing the training and validation losses, alongside a qualitative judgment to discard overly complicated expressions.

When considering galaxy properties, our input feature set consists of
$\log_{10} (M_\star/{\rm M_{\odot}})$, 
$\log_{10}(M_{\rm gas}/{\rm M_{\odot}})$, 
$\log_{10} (M_{\rm dust}/{\rm M_{\odot}})$,
${\rm SFR}/{\rm M_{\odot} yr^{-1}}$, 
$Z_{\rm g}$, 
$r_{\star}/\rm{kpc}$, 
$r_{\rm g}/\rm{kpc}$, 
$\Sigma_{\rm SFR}/{\rm M_{\odot} yr^{-1} \rm{kpc^{-2}}}$, and 
$\sin \theta_{\rm inc}$ (exact definitions of these quantities are provided in Tab. 4 in \citealt{Sommovigo_2025}). 
For ${\rm SFR}$ we use the values averaged over $100 \, {\rm Myr}$, i.e. ${\rm SFR} = {\rm SFR}_{100}$ and ${\rm sSFR} = {\rm sSFR}_{100}$.
In addition, we include a dust-specific parameter, $\xi$, defined as the dust mass fraction contained in grains with radii $a < 1700 \, \angstrom$. The values of $\xi$ for the different dust mixtures are:
\begin{equation}
    \xi_{\rm MW} = 0.5968, \;
    \xi_{\rm SMC} = 0.7193, \;
    \xi_{\rm stellar} = 0.3064.
\end{equation}

We fit a single symbolic-regression model across all dust mixture models for $\Av$, $B_1$, and $B_3$. In practice, the resulting expressions retain some dependence on $\xi$, reflecting systematic differences between the dust mixtures.
For $B_0$ and $B_2$ we explored two approaches: fitting separate functions for each dust mixture, and fitting a single unified function in which the dust mixture dependence is encoded through the parameter $\xi$.
We found that better predictions were made for $B_0$ if a different function for each dust mixture was found, which we attribute to the very different distributions of parameters for each dust mixture, as illustrated in \cref{fig:corner_allpar_all_dust}.
For $B_2$, we were able to obtain a single expression which encapsulated the behavior of all dust mixtures.

\subsubsection{Learned analytic expressions}

For our first equation, we consider $\Av$ as a function of galactic properties. Upon running \operon, we find that $\log_{10}\Av$ can be approximated as
{
\begin{equation}
    \label{eq:Av_fit}
    \begin{split}
         \log_{10}\Av & \approx \alpha_0 - \frac{\alpha_1}{\log_{10}
         (\alpha_2 \sin \theta_{\rm inc})} - \frac{\alpha_3}{\log_{10}(\alpha_4 \xi)} \\
         & - \alpha_5 (\alpha_6 \Sigma_{\rm SFR})^{-\alpha_7 Z_{\rm g}},
    \end{split}
\end{equation}}
where $\alpha = \{\splitatcommas{0.428, 0.00967, 0.953, 0.00383, 1.51, 1.68, 800 \, ({\rm M_{\odot} yr^{-1} \rm{kpc^{-2}}})^{-1} , 4.75}\}$,
giving a $R^2$ on the training and validation set of 0.615 and 0.650, respectively.
We note that the equation returned by \operon{} had a numerator of $\alpha_1 \log_{10}(M_{\rm dust}/{\rm M_\odot})$ instead of $\alpha_1$ (with a different optimized value of $\alpha_1$).
However, $M_{\rm dust}$ is only constrained in observations if one has access to the rest-frame FIR emission, and thus that formula would not be applicable for galaxies that are only detected in the UV, although it could prove useful for forward modeling approaches, since dust mass is available in simulations (or can be approximated based on gas-phase metallicity).
We find that this replacement does not significantly reduce the value of $R^2$, hence we prefer the expression given in \cref{eq:Av_fit}.

The recovered expression is physically interpretable and remarkably consistent (in terms of the global galaxy properties involved) with the analytical attenuation model proposed in \citet[][Eq.~5]{Sommovigo_2025}. In particular, \operon{} independently recovers mostly the same key quantities previously identified as the main drivers of V-band attenuation: viewing angle, SFR surface density, metallicity, and, in addition, dust composition.

The $\sin \theta_{\rm inc}$ term increases monotonically from face-on to edge-on configurations, reflecting the longer path length through the galactic dust disk at higher inclination\footnote{This dependence plays a role analogous to the geometrical factor $f_\mu=\cos\theta$ introduced in \citet{Sommovigo_2025}}.
The coupled $\Sigma_{\rm SFR}$--$Z_{\rm g}$ term captures the effective dust surface density in star-forming gas. As discussed in \citet{Sommovigo_2025}, the dependence on $\Sigma_{\rm SFR}$ naturally emerges from connecting the gas surface density to star formation through the Kennicutt--Schmidt relation \citep{Kennicutt98}, while $Z_{\rm g}$ regulates the dust-to-gas ratio through the adopted metallicity-dependent dust scaling \citep{RemyRuyer14}. Their coupled appearance therefore reflects the fact that the optical depth depends both on the star forming gas surface density and on how dust-enriched that gas is. As $\Sigma_{\rm SFR}$ and/or $Z_{\rm g}$ increase, 
the final term of \cref{eq:Av_fit}
asymptotically approaches zero, causing $\log_{10} \Av$ to saturate toward large attenuation values.
Finally, the dust-composition-related parameter $\xi$, which we did not explicitly probe in \citet{Sommovigo_2025}, enters with a comparatively small coefficient ($\alpha_3 \approx 0.004$), indicating that grain properties play a secondary role for $\Av$ relative to geometry and dust column density. This is physically expected, since the normalization of the attenuation curve is primarily regulated by the total dust mass along the line of sight rather than by the detailed grain population.

For our next two parameters, $B_1$ and $B_3$, we obtain functions of both galactic properties as well as $\Av$. After analyzing the \operon{} outputs, we choose as our learned analytic expressions
{
\begin{equation}
    \label{eq:B1_fit}
    B_{1 {\rm s}} = \frac{\alpha_0}{\Av}
    \left[ 
    (\alpha_1 Z_{\rm g})^{\alpha_2 \sin \theta_{\rm inc}} - \alpha_3 \frac{\log_{10} (M_\star / {\rm M_\odot})}{\log_{10} (\alpha_4 \xi)}
    \right],
\end{equation}
}where $\alpha = \{ \splitatcommas{0.0324, 25.5, 2.36, 0.00411, 1.95} \}$, 
and
{\begin{equation}
    \label{eq:B3_fit}
    \begin{split}
         B_3 \approx & - \alpha_0 - \alpha_1 \sin \theta_{\rm inc} + \alpha_2 Z_{\rm g} (\alpha_3 \Av)^{-\alpha_4 {\rm sSFR}} \\
         & + \alpha_5 (\alpha_6 \Av)^{-\alpha_7 \xi},
    \end{split}
\end{equation}
}with $\alpha = \{ \splitatcommas{2.1, 0.83, 51.1, 21.2, 9.7 \, {\rm Gyr}, 3.52, 0.0417, 0.17} \}$. 
These gave a $R^2$ on the training and validation set of 0.537 and 0.529, respectively, for $B_1$, whereas for $B_3$ these are 0.692 and 0.750.

For $B_1$, which modulates the FUV slope, the $1/\Av$ prefactor implies that deviations in the FUV slope are largest at low optical depth and progressively suppressed as $\Av$ increases. This behavior is physically expected: once the effective optical depth becomes large, additional increases in dust column produce progressively smaller changes in the transmitted spectrum, since the attenuation asymptotically saturates at high optical depth. Although an exact exponential attenuation law applies only in the idealized foreground-screen limit, the same qualitative behavior is expected more generally in radiative-transfer calculations. The remaining dependence on $Z_{\rm g}$, $\sin \theta_{\rm inc}$, and $M_\star$ captures secondary modulation of the UV slope by dust composition, viewing geometry, and global galaxy mass budget. Interestingly, the appearance of $M_\star$ is consistent with the mutual-information analysis, where stellar mass retained significant predictive power despite exhibiting only weak Spearman correlations \citep[see also][]{Salim18}.

For $B_3$, which controls the large-scale UV-to-optical slope, the recovered expression reproduces the well-known tendency for heavily obscured sightlines to exhibit grayer attenuation laws. The inclination enters with a negative coefficient ($-0.83\sin \theta_{\rm inc}$), so that closer to edge-on sightlines yield lower $B_3$ (grayer/flatter curves), consistent with the higher dust column density along the los at high inclination.  $Z_{\rm g}$ and $\xi$ act as amplifiers of the slope–attenuation coupling: higher metallicity increases the overall dust content at fixed gas mass, while larger values of $\xi$ steepen the dependence on $\Av$ through the exponent of the second power-law term, enhancing the intrinsic UV-to-optical opacity contrast by increasing the abundance of small grains. The sSFR appears in the exponent of $\Av$: at fixed $\Av$ and $Z_{\rm g}$, galaxies with higher specific star-formation rate see the $Z_{\rm g}$-weighted term suppressed, reflecting the fact that actively star-forming systems host younger, more UV-luminous stellar populations whose radiation is preferentially processed by the surrounding dust, modifying the effective attenuation slope. The fact that five input quantities ($\Av$, $Z_{\rm g}$, $\xi$, $\sin \theta_{\rm inc}$, and sSFR) are required to reproduce $B_3$ indicates that the large-scale attenuation slope is regulated not only by dust column density and grain composition, but also by the star–dust geometry (via $\sin \theta_{\rm inc}$) and the strength of the radiation field (via sSFR).

Our learned analytic expression for $B_0$ was obtained by considering functions of $\Av$, $B_1$ and $B_3$. Given that this parameter controls the strength of the bump, and that only the MW dust mixture contains this feature, we find that the distribution of $B_0$ values is very different for MW compared to the SMC and stellar dust mixtures. As such, when we attempted to find a single function for all dust mixtures, the expressions contained highly complex functions of $\xi$ which acted like a delta-function for the MW value. To avoid this presumably fine-tuned behavior, we instead chose to find a different function for the MW dust mixture and for SMC and stellar mixtures. For the MW, we obtained the learned analytic expression
{
\begin{equation}
    \label{eq:B0_fit_MW}
    \begin{split}
        \log_{10} B_0 &\approx \alpha_0 B_{1, {\rm s}} + \alpha_1 B_3 \\
        & - \frac{\alpha_2}{B_3} \left( \alpha_3 B_{1, {\rm s}} + \alpha_4 B_3 + (\alpha_5 \Av)^{\alpha_6 B_3} \right),
    \end{split}
\end{equation}
}where the coefficients were optimized to only consider the values of $\log_{10}B_0 > -0.5$ to remove outliers. These optimized coefficients are
$\alpha = \{ \splitatcommas{0.662, 0.224, 0.327, 13.8, 1.53, 23.7, 0.112} \}$,
which yielded $R^2$ values of 0.931 and 0.927 for the training and validation sets, respectively, after applying this cut on the true value of $B_0$.
These large values of $R^2$ are consistent with the internal parameter correlations presented in \cref{fig:internal_corr_dust}, where we found that $B_0$ and $B_3$ were highly correlated for the MW dust mixture, with a Spearman rank coefficient of 0.92.

The recovered dependence on both $\xi$ and on attenuation/slope-related quantities ($\Av$ and $B_3$) is physically reasonable. Larger values of $\xi$, corresponding to a higher fraction of small grains carrying the UV bump feature, naturally lead to stronger bumps. Conversely, the suppression of $B_0$ toward flatter attenuation curves and larger optical depths is consistent with the well-known tendency for the UV bump to weaken in highly obscured sightlines \citep{Narayanan18, Matsumoto26}. Physically, at large optical depth the emergent UV spectrum becomes increasingly dominated by leakage through low-optical-depth paths and by scattering effects, both of which tend to dilute narrow spectral features such as the UV bump. We defer a more detailed comparison with previous literature results to the next section.

\begin{figure*}
    \centering
    \includegraphics[width=0.95\textwidth]{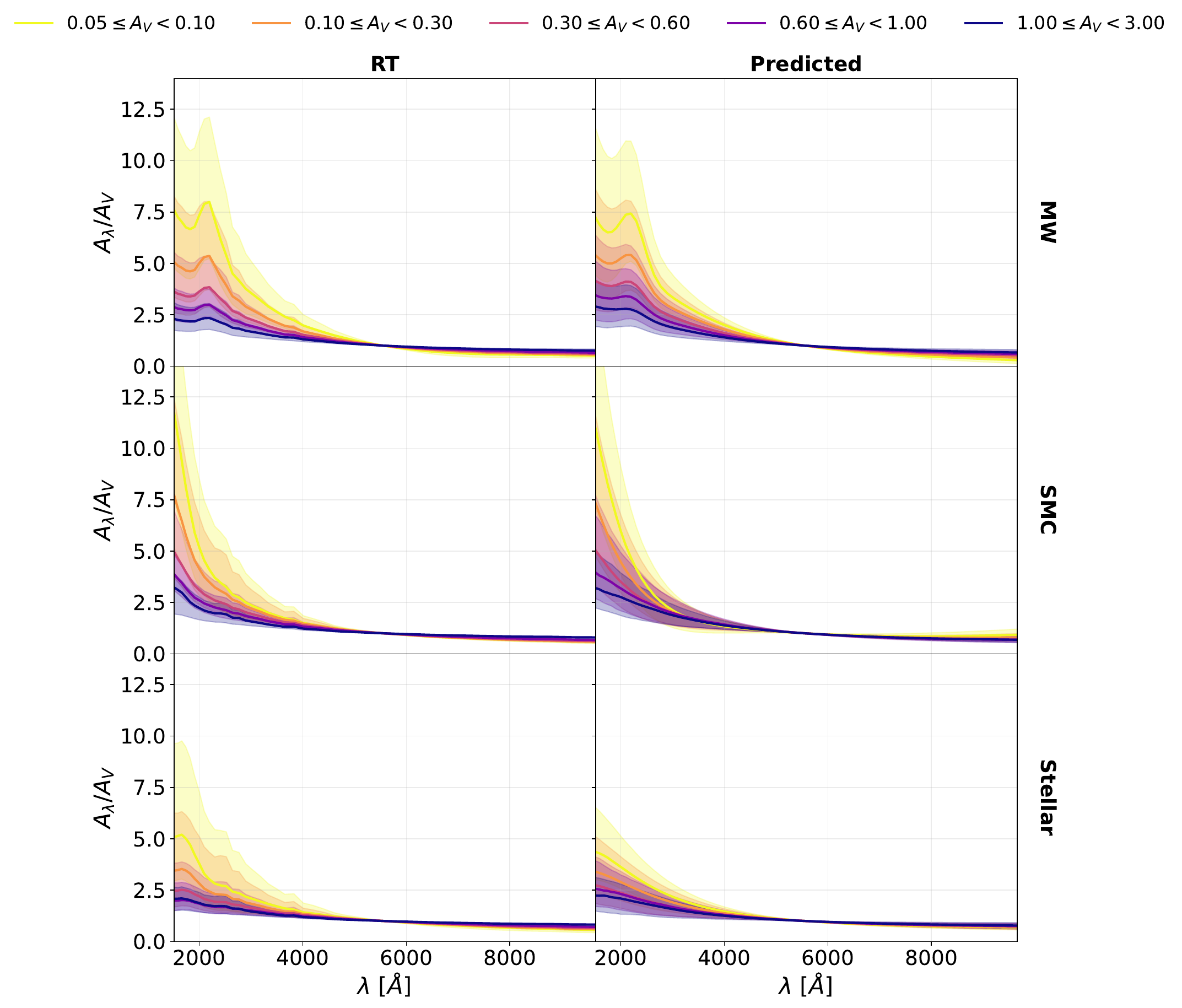}
    \caption{Distribution of attenuation curves from radiative transfer (left) or from our learned analytic expressions for the dust attenuation curve parameters (right) for different dust mixtures, as a function of $\Av$. We plot the median and 68\% distribution of values at each wavelength. The distributions are similar for the true and generated curves, for every dust mixture.
    }
    \label{fig:predicted_curves}
\end{figure*}

For the combined stellar and SMC dust mixtures, our learned analytic expression for $B_0$ is:
{
\begin{equation}
    \label{eq:B0_fit_SMC_stellar}
    \begin{split}
         \log_{10} B_0 & \approx \alpha_0 B_3 [ 
        \alpha_1 \Av + (\alpha_2 \Av)^{\alpha_3 B_{1, {\rm s}}} \\
        & - \alpha_4 \log_{10}(\alpha_5 B_3) + (\alpha_6 \xi)^{-\alpha_7 B_3}
       ]
        - \alpha_8.
    \end{split}
\end{equation}
}To avoid outliers, we optimized the parameters of this expression by removing all values of $\log_{10}B_0$ greater than $-2$, and obtained
 $\alpha = \{ \splitatcommas{0.0413, 5.79, 10.2, 3.78, 5.13, 7.20, 0.379, 0.127, 2.91} \}$. On the cut samples, this gave a $R^2$ of 0.712 and 0.632 for training and validation, respectively. The lower values of $R^2$ compared to the MW are understandable given \cref{fig:internal_corr_dust}, where one observes smaller internal correlations between $B_0$ and the other parameters for the SMC and stellar dust mixtures.

Unlike $B_0$, we found that a single expression could be obtained to approximate $B_2$ as a function of the other attenuation curve parameters and $\xi$. This was found to be:
{
\begin{equation}
    \label{eq:B2_fit}
    \begin{split}
        B_{2, {\rm s}} & \approx \alpha_0 B_0 + \alpha_1 \\
        & - \xi \left[ \alpha_2 B_0^{-\alpha_3} + \alpha_4 B_{1, {\rm s}} \left( \alpha_5 B_{1, {\rm s}} + \log_{10}(\alpha_6 B_0) \right)\right],
    \end{split}
\end{equation}
}where $\alpha = \{ \splitatcommas{0.264, 0.129, 0.00351, 0.776, 1.29, 5.93, 179} \}$.
This function is able to predict $B_2$ well, with $R^2$ values of 0.743 and 0.809 for our training and validation samples, respectively.

\subsubsection{Mock attenuation curves}

We now assess the efficacy of our learned equations for the attenuation curve parameters in reproducing realistic attenuation curves. We consider all galaxies and lines of sight within each sample, and predict a value of $\Av$, $B_0$, $B_1$, $B_2$ and $B_3$. For each variable, we first compute the mean prediction based on the galaxy properties and/or the other (predicted) attenuation curve parameters, then add Gaussian noise of magnitude
$\sigma(\log_{10} \Av) = 0.221$,
$\sigma(\log_{10} B_0) = 0.072$ for MW dust and $\sigma(\log_{10} B_0) = 0.216$ for SMC/stellar dust,
$\sigma(B_{1, {\rm s}}) = 0.058$,
$\sigma(B_{2, {\rm s}}) = 0.381$, and
$\sigma(B_3) = 0.728$.
These values are calibrated by considering the residuals on the fits in our training set.
To prevent unrealistic values of these parameters, we then clip their values to the range
$\log_{10} \Av \in [-5, 2]$,
$\log_{10}B_0 \in [-5.0, 2]$,
$B_{1, {\rm s}} \in [-1, 1]$,
$B_{2, {\rm s}} \in [-5, 5]$, and
$B_{3} \in [0.03, 10]$.
We note that variables which depend on other attenuation curve parameters are computed using the noised and clipped values of the parameters, as opposed to either their true values or the mean predictions from our formulae.

The distributions of resulting attenuation curves are plotted alongside those obtained from radiative transfer in \cref{fig:predicted_curves}, where we separate the samples by the value of $\Av$. The qualitative trends as dust mixture or $\Av$ changes are the same, whether considering the generated or true attenuation curves. The mean predictions for each $\Av$ bin and the scatter around these are similar for the two cases, indicating that this procedure produces realistic looking attenuation curves and thus could be used to produce mock photometric observations from synthetic galaxy catalogs.
In the FUV, for the stellar dust model, the predicted attenuation curves do not flatten to the same extent as those obtained through RT. Since this behavior is determined by $B_1$, for which we obtained a single function for all dust mixtures, perhaps promoting this to a different function for each dust mixture could alleviate such differences. We leave a more detailed study of the properties of such synthetic attenuation curves in generating mock galaxy catalogs to future work.

\section{Discussion}\label{sec:discussion}

In this section, we briefly place our results in the context of previous observational and theoretical studies of dust attenuation curves, and discuss the main caveats of the present analysis. Overall, our results support a picture in which the large-scale normalization and slope of attenuation curves are primarily regulated by galaxy-scale properties and viewing angle, while the detailed UV structure retains a stronger sensitivity to the underlying dust composition and dust and stellar spatial distributions.

\subsection{Comparison with previous theoretical and observational studies}

Our results broadly support the now well-established picture that the large-scale shape of dust attenuation curves is primarily regulated by the interplay between dust column density, viewing geometry, and radiative-transfer effects \citep{Chevallard13,Narayanan18,Salim18,Salim20,Trayford20,Witt00}. We recover the canonical correlation between attenuation curve slope and $\Av$ -- with flatter, grayer curves at higher optical depths -- which has been reported in essentially every observational and simulation study to date \citep[e.g.][]{Witt00,Kriek13,Salim18,Narayanan18,Trayford20}. More specifically, our symbolic-regression expression for the large-scale slope parameter $B_3$ (\cref{eq:B3_fit}) explicitly contains a $\log_{10}(\alpha_2 \Av)$ dependence, making this attenuation--slope coupling a quantitative, testable prediction.

A more specific finding of our analysis is that $\Sigma_{\rm SFR}$, particularly normalized to the young stellar half-mass radius, emerges as the single strongest galaxy-property predictor of $\Av$ ($r \sim 0.5$--$0.6$), surpassing both stellar mass and metallicity in raw correlation strength. 
This is consistent with previous RT studies of EAGLE, MUFASA, and FirstLight galaxies \citep{Narayanan18,Trayford20,Mushtaq2023}, and with observational studies finding strong sSFR/$\Sigma_{\rm SFR}$ dependence in local and intermediate-redshift samples \citep{Reddy15,Battisti2016,Battisti20,Salim18,Shivaei20,Maheson25}. The fact that $\Sigma_{\rm SFR}$ rather than total SFR, $Z$, $M_{\rm gas}$, or $M_\star$ emerges as the cleanest predictor supports a picture in which the effective optical depth along a sightline is set jointly by the intensity of the UV radiation field and the spatial concentration of its sources relative to the dust distribution.

The symbolic-regression expression we recover for $\Av$ (\cref{eq:Av_fit}) provides an independent, data-driven verification of the analytical model proposed in \citet{Sommovigo_2025}: \operon{} recovers the same dominant variables ($\sin \theta_{\rm inc}$, $\Sigma_{\rm SFR}$, $Z_{\rm g}$), in qualitatively the same functional roles --- viewing geometry, star-forming gas surface density, and metallicity-dependent dust enrichment. The agreement is non-trivial, since the SR had access to a wider set of galaxy properties and was not biased toward the \citet{Sommovigo_2025} form. The principal addition from our SR analysis is the explicit appearance of a dust-composition parameter $\xi$ (the small-grain mass fraction), entering with a small coefficient ($\alpha_3 \sim 0.004$), confirming that $\Av$ is primarily set by geometry and column density and only secondarily by grain population.

Our results also support the growing evidence that variations in attenuation curve shape cannot be interpreted purely as a signature of dust composition, but instead emerge from the coupling between grain properties and star--dust geometry \citep{Witt00,SeonDraine16,Narayanan18,Lin21,Vijayan24,Sommovigo_2025,Matsumoto26}. In agreement with these works, we find that the $2175\,\angstrom$ bump becomes progressively weaker for flatter and more highly attenuated sightlines: our MW-only SR expression for the bump strength $B_0$ (\cref{eq:B0_fit_MW}) explicitly encodes the bump suppression with increasing $\Av$ and increasing $B_3$ (flatter slope), reaching $R^2 \sim 0.93$ on the training set. $\Av$, the small-grain fraction $\xi$, and the clumpiness of the star--dust distribution all act as bump regulators, in agreement with the qualitative pictures of \citet{Narayanan18} and \citet{Matsumoto26}, underscoring the difficulty highlighted by \citet{Lin21} of constraining microscopic grain properties from integrated galaxy SEDs. The role of star--dust clumpiness, which requires higher-resolution simulations resolving the multiphase ISM, will be explored in follow-up work.%

A further point of contact with the literature concerns the role of stellar mass, which behaves quite differently in linear versus nonlinear statistical measures. While $M_\star$ shows only weak Spearman correlations with attenuation curve shape parameters in our analysis, it gains substantial predictive power in the mutual-information matrix and appears explicitly in our SR expression for $B_1$ (\cref{eq:B1_fit}). The same is true of other macroscopic properties such as gas mass $M_{\rm g}$, gas metallicity $Z_{\rm g}$, and characteristic stellar/gas sizes, which retain residual mutual information with $B_0$ and $B_2$ even when monotonic correlations vanish in the combined dust mixture sample. This is consistent with \citet{Salim18}, who found that more massive (and more optically thick) galaxies systematically display flatter curves, and with the recent intermediate-redshift analysis of \citet{Maheson25} reporting similar mass-dependent trends. More generally, $M_\star$ acts as a proxy for the combined effects of metallicity, dust mass, and structural compactness; our analysis suggests that part of the apparent weakness of the $M_\star$--slope correlation in single-dust mixture studies arises from its non-monotonic, nonlinear character, which Spearman-based statistics underestimate. This also indicates that the same galaxy-scale properties continue to regulate UV attenuation across dust mixtures, but with detailed mappings onto UV slope and bump strength that are increasingly dust mixture dependent.

Finally, our results are directly relevant in the context of recent JWST observations of galaxies at $z\gtrsim4$, which increasingly report flat, often featureless attenuation curves \citep{Witstok23,Markov24,Fisher25,Lin25,Ormerod25,Shivaei25,Chworowsky26,Rodighiero26}. 
Several studies have interpreted these trends as evidence for evolving grain populations, with larger grains produced by stellar sources dominating at early times and small grains appearing later through ISM processing \citep{Markov24,McKinney25,Narayanan25,Trayford26}. Our results are consistent with this interpretation in that the small-grain fraction $\xi$ enters explicitly in our SR expressions for $\Av$, $B_3$, and $B_0$, controlling precisely the UV-sensitive aspects of the attenuation curve. However, the strong line-of-sight and geometry-driven scatter identified here \citep[see also][]{Sommovigo_2025,Sommovigo26}, combined with the fact that very different grain populations can produce nearly indistinguishable featureless curves -- as is the case for our SMC and stellar dust mixtures at high $\Av$ (see \cref{fig:predicted_curves}) -- highlights the difficulty of uniquely inferring grain properties from integrated attenuation curves alone, particularly at the high $\Av$ end and in unresolved high-redshift observations. This is especially relevant for current JWST analyses that adopt low dimensional parametric forms and attempt to discriminate between SMC-like (small-grain-rich, low-metallicity) and large-grain-dominated (stellar-injected) dust populations from the inferred slope alone. Disentangling these scenarios likely requires either spatially resolved observations or independent constraints on the dust and stellar surface density.

Beyond reproducing previously identified trends, the SR framework introduced here provides a quantitative decomposition of how geometry, optical depth, dust column density, and grain composition jointly shape attenuation curves. The resulting analytic expressions can be directly incorporated 
into 
numerical galaxy formation simulations, including semi-analytic and hydrodynamic models,
to produce synthetic galaxy catalogs, or SED fitting procedures without requiring full radiative-transfer post-processing.

\subsection{Caveats and limitations of the present analysis}\label{sec:caveats}

While the framework presented in this work provides a systematic, quantitative decomposition of the dust attenuation curve and its dependence on galaxy properties and dust composition, several caveats and limitations should be kept in mind.

First, our analysis is based on a single set of synthetic attenuation curves derived from \skirt{} post-processing of TNG50 and TNG100 local galaxies, from a single snapshot (snapshot 93, $z=0.07$). The resolution and sub-grid physics of these simulations -- in particular the smoothed, effectively single-phase ISM treatment -- limit how faithfully the small-scale structure of the dust and its relative geometry with respect to young stars is captured, as discussed also in \citet{Sommovigo_2025}. Higher-resolution zoom-in simulations resolving a genuinely multiphase ISM, and incorporating birth-cloud and GMC-scale structure, can produce systematically different attenuation curves at fixed global galaxy properties \citep{Choban2022,Choban2024,DiMascia24,Dubois24,Narayanan25,Matsumoto26,Trayford26}. It is therefore important to test whether the functional form and the scaling relations derived here remain optimal when applied to attenuation curves extracted from such simulations -- in particular those resolving the optical-depth structure of young stellar birth clouds, which play a central role in shaping the FUV slope \citep{CharlotFall2000,Matsumoto26}. We defer this to future work.

A related limitation concerns applicability at higher redshift. Since our library is constructed from local galaxies, the scaling relations linking attenuation curve parameters to galaxy properties should be extrapolated with caution to systems with substantially different star formation rate, compactness, metallicities, and dust content. We expect the qualitative trends to persist, but not necessarily the precise coefficients. By contrast, the proposed four-parameter functional form itself is likely more robust, as it captures a broader range of attenuation curve shapes than typically inferred from SED fitting across both local and high-redshift galaxy samples \citep{Salim18,Markov24,Fisher25,Shivaei25,Chworowsky26,Rodighiero26}.

Second, our parameterization was selected by optimizing reconstruction accuracy on synthetic attenuation curves, where the underlying ``true'' attenuation curve is known by construction. In practice, attenuation curves are inferred from observed SEDs through Bayesian SED fitting, where the relevant question is not whether a more flexible parameterization can fit synthetic curves more accurately, but whether the additional flexibility is actually \textit{warranted} by the observational data given finite photometric coverage and signal-to-noise. 
This is not straightforward to assess: Bayesian model-comparison metrics (e.g.\ the evidence, or information-criterion-based estimators) applied to real SEDs are sensitive to degeneracies with other model components (in particular the star formation history and metallicity) where the degree of complexity of their distributions is not known \textit{a priori}, making it difficult to isolate the contribution of the attenuation parameterization. In this context, our conclusion that attenuation curves are intrinsically four-dimensional provides a physically motivated prior on the \textit{expected} complexity of the attenuation model, independent of these degeneracies, and can help break the circularity of choosing a parameterization before fitting. A natural next step is to incorporate this prior into Bayesian SED fitting and assess whether the four-parameter form derived here is preferred over simpler parameterizations such as those of \citet{Calzetti00} or \citet{Salim18} and \citet{Noll2009}.

Third, although we attempt to disentangle the effects of dust mixture, galaxy properties, and radiative-transfer geometry, a fully systematic separation would require a substantially larger synthetic library spanning a broader range of simulations and dust models. In particular, realistic galaxies likely populate a continuous distribution of grain size distributions and compositions that evolves with metallicity and star-formation history \citep{Asano13,Hirashita19,Aoyama:2020,Choban2022,Choban2024,Dubois24,Narayanan25,Montero26,Trayford26}. In this work, the dust mixture is treated as an independent, discretized  
axis, meaning that potential correlations between galaxy properties and dust composition are not captured. Coupling the dust mixture self-consistently to galaxy evolution through live-dust models would address this limitation, but at the cost of introducing additional model dependence and reducing our ability to isolate dust mixture effects independently.

Finally, SR is a stochastic optimization procedure: \operon{} explores a vast space of analytic expressions guided by a multi-objective Pareto front, and different runs can converge to functional forms that are statistically comparable in $R^2$ but structurally distinct. The expressions presented here should therefore be interpreted as one representative solution from a family of plausible analytic descriptions, rather than as a uniquely determined functional form. The recovered form may be somewhat specific to the simulation setup used to generate the training library and could differ if alternative simulation suites with different ISM/dust prescriptions were used. 

Taken together, these caveats define a natural path forward: extending the synthetic library to higher-resolution simulations with resolved multiphase ISM and dynamical dust models, validating the parameterization against real observational data through Bayesian model selection, and quantifying the robustness of the symbolic-regression results across independent simulation suites.

\section{Summary}\label{sec:conclusions}

Dust attenuation remains one of the main sources of uncertainty in both SED fitting and forward modeling of synthetic observations, in part because widely used attenuation laws are empirical, low-dimensional, and not obviously tied to the physical properties of galaxies. In this paper, we used attenuation curves derived from \skirt{} radiative-transfer calculations for TNG50 and TNG100 galaxies, each viewed along multiple lines of sight and
post-processed with three dust mixtures (MW, SMC, and stellar dust), to construct a more physically motivated description of attenuation curve space. We employed an Information-Ordered Bottleneck autoencoder to characterize the effective dimensionality of the curves, symbolic regression to derive a new interpretable functional form, and a second
symbolic-regression step to connect the inferred attenuation curve parameters to galaxy properties. This framework allowed us both to assess the limitations of commonly used parameterizations and to derive
predictive relations that can be used when full radiative transfer is impractical.

Our main conclusions are as follows.

\begin{itemize}
    \item To fully describe the diversity of possible attenuation curves, our IOB analysis indicated a minimum requirement of four free parameters. The intrinsic dimensionality is reduced for some dust mixtures, most clearly for bump-less curves when the dust mixture does not contain carbonaceous grains. Any lower-dimensional parameterization may 
    perform adequately in restricted regimes, but it does not provide a sufficiently general description of the full range of curves.

    \item We derived a new attenuation curve model, given by \cref{eq:final_model}, with four free parameters,
    $\{B_i\}$. This parameterization is physically interpretable: $B_0$ controls the strength of the UV bump, $B_1$ controls the FUV slope, $B_2$ controls the curvature around the UV-bump transition region, and $B_3$ controls the large-scale slope and long-wavelength limit of the attenuation curve. Compared with commonly used models from the literature, we
    find that this new functional form provides a significantly better fit to the full range of attenuation curves considered here, reducing the error on the recovered intrinsic (unattenuated) flux down to $<10\%$.

    \item Internal correlations between parameters are reduced in the new parameterization compared to traditional models, making the fits more stable. At the same time, the correlation structure changes with dust mixture, so the
    dimensionality cannot be reduced in a universal way unless one restricts the analysis to a single dust mixture. Different dust mixtures occupy different regions of parameter space, potentially allowing the dust composition to be constrained from inferred attenuation curve parameters.

    \item We characterized the correlations between attenuation curve parameters and galaxy properties using both Spearman and mutual-information statistics, and derived approximate analytic expressions for each parameter via symbolic regression (\cref{eq:Av_fit,eq:B1_fit,eq:B2_fit,eq:B0_fit_MW,eq:B0_fit_SMC_stellar,eq:B3_fit}). The recovered trends reveal that attenuation curve diversity is jointly regulated by the underlying dust composition (small-grain fraction $\xi$, dust mixture), galaxy-scale physical conditions (most strongly $\Sigma_{\rm SFR}$ and metallicity), and radiative-transfer geometry (inclination and line-of-sight effective optical depth). These dependencies are robust across all three dust mixtures considered, and our recovered $\Av$ expression mirrors the physically motivated analytical model of \citet{Sommovigo_2025}, despite being derived from a wider input feature set. Beyond the dominant trends, properties such as $M_\star$, $M_{\rm g}$, and characteristic sizes retain substantial \textit{nonlinear} predictive power that linear correlation statistics underestimate. The resulting expressions can be used in forward modeling or to define physically motivated priors for SED fitting.

\end{itemize}

Taken together, these results show that the commonly adopted attenuation curve prescriptions are not flexible enough to describe the full diversity of attenuation curves present in modern radiative-transfer
galaxy simulations, and that our new four-parameter model provides a substantially better description of that space. Within the scope tested here, our model should therefore be preferred over the standard prescriptions
in applications that require parametric attenuation curves. The derived scaling relations further make it possible to assign physically motivated attenuation curves in SED fitting, forward modeling, and mock-catalog generation even when full radiative-transfer calculations are not feasible.

At the same time, this conclusion is currently calibrated only on TNG galaxies at a single redshift, with a fixed dust-to-metal ratio and a limited, albeit representative, set of dust mixtures. It remains to be
tested how broadly the parameterization and the inferred scaling relations generalize to other simulation suites, evolving galaxy populations, and real observational datasets, and whether the same functional form remains preferred once model evidence is taken into account. This should be the subject of future work, and it is possible that further refinements to the functional form may be needed to capture the full diversity of attenuation curves across different regimes. However, the framework developed here provides a clear path for such future investigations, 
and the results obtained so far suggest that a more flexible, physically motivated parameterization can significantly improve our ability to model dust attenuation in galaxies.

More broadly, this work shows how simulations can be used to derive analytic and interpretable prescriptions that are accurate enough for practical use. The same methodology developed here --- combining dimensionality estimation, symbolic regression, and validation against full radiative-transfer calculations --- can be applied more generally to derive empirical analytic models from simulations in other areas where direct numerical calculations remain too expensive for routine inference and forward modeling.

\section*{Acknowledgments}
We thank Kartheik Iyer and Shy Genel for insightful suggestions on the project and Aaron Yung for useful comments and suggestions on the manuscript.
The Flatiron Institute is supported by the Simons Foundation. 
This project was developed as part of the Simons Collaboration on ``Learning the Universe.'' 
DJB acknowledges that support was provided by Schmidt Sciences, LLC.
RKC is grateful for support from the Leverhulme Trust via the Leverhulme Early Career Fellowship. 
CCL was supported by the research environment and infrastructure of the Handley Lab at the University of Cambridge.
We thank Jonathan Patterson for smoothly running the Glamdring Cluster hosted by the University of Oxford, where some of the data processing was performed. The radiative transfer simulations and analyses presented in this work were run on the Flatiron Institute’s research computing facilities (Popeye).

\appendix

\begin{figure*}[t]
\includegraphics[width=\linewidth]{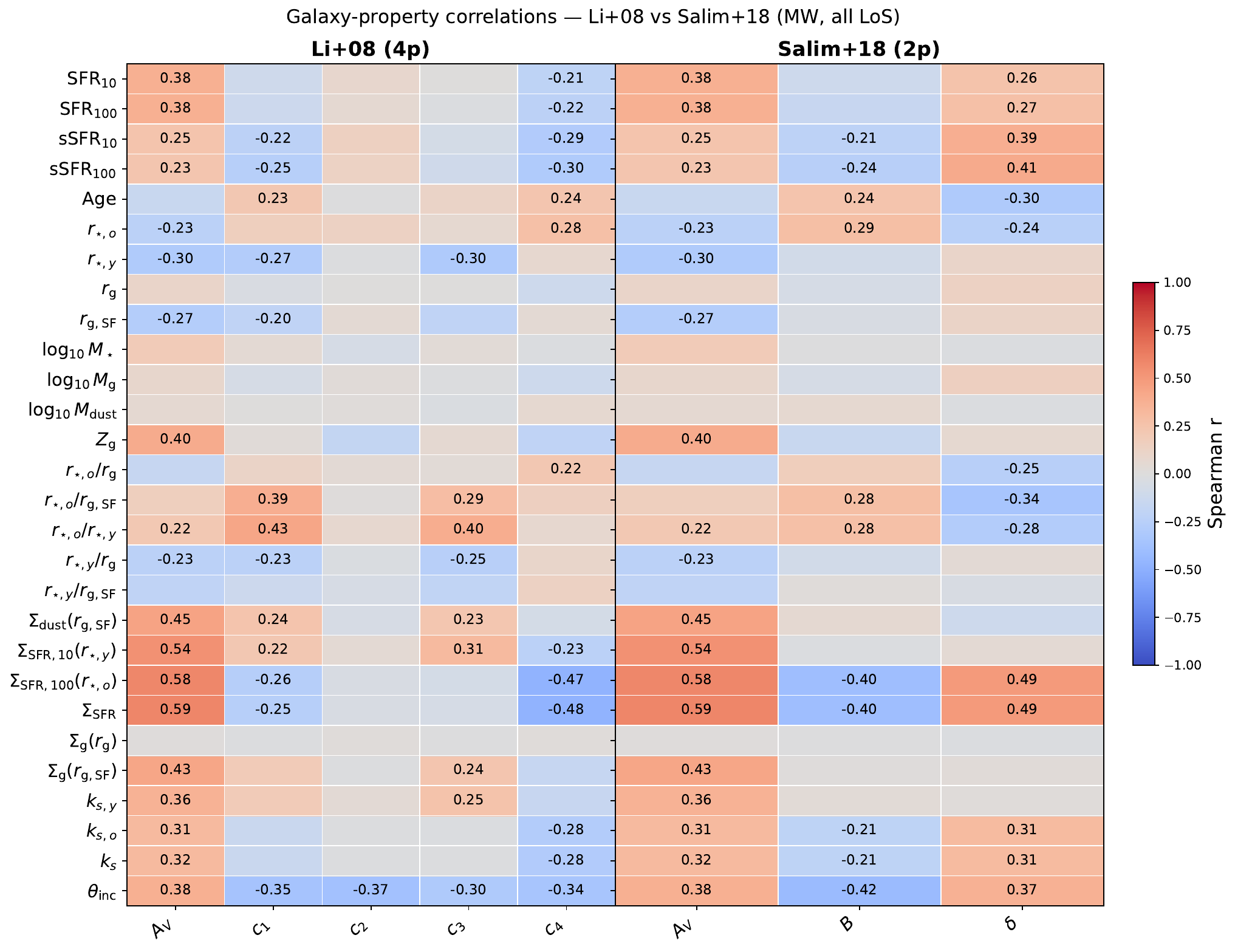}

\caption{
Correlation matrices between attenuation curve parameters and galaxy properties for the \citet{Li08} and \citet{Salim18} parameterizations, computed using all lines of sight and assuming Milky Way dust. 
Values of the Spearman correlation coefficient, $r$, obeying $r < 0.2$ are not printed for clarity.
Compared to our new parameterization, these models exhibit a lower fraction of significant correlations and a less structured mapping onto galaxy properties, highlighting the improved interpretability and physical connection enabled by our functional form.
}
\label{fig:correlations_other_models_MW}
\end{figure*}

In this Appendix, we consider the correlations between the inferred parameters of dust attenuation models and the galaxy properties defined in Section~\ref{sec:galaxy_correl}.
\cref{fig:correlations_other_models_MW} gives the Spearman correlation coefficients between these properties and the parameters of the \citet{Li08} and \citet{Salim18} models for the MW dust mixture. These correlations should be compared to \cref{fig:correlations_newmodel_dust_combined}, where we show the analogous results for the model introduced in this paper.

\begin{figure*}[t]
\includegraphics[width=\linewidth]{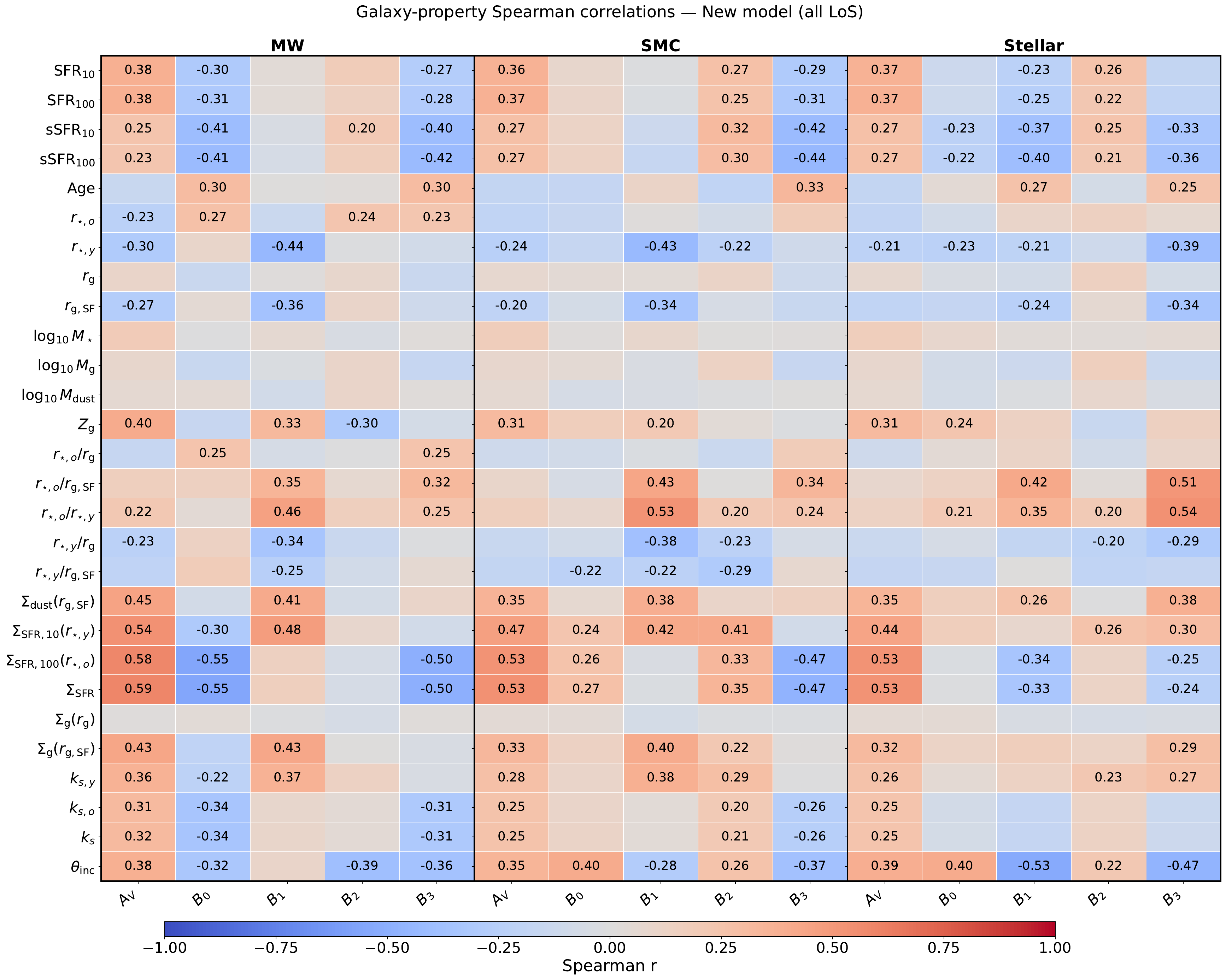}

\caption{Same as Fig.~\ref{fig:correlations_newmodel_dust_combined} but for individual dust mixtures (MW, SMC, and stellar from left to right) separately.
}
\label{fig:correlations_dust_mixes}
\end{figure*}

\begin{figure*}[t]
\includegraphics[width=\linewidth]{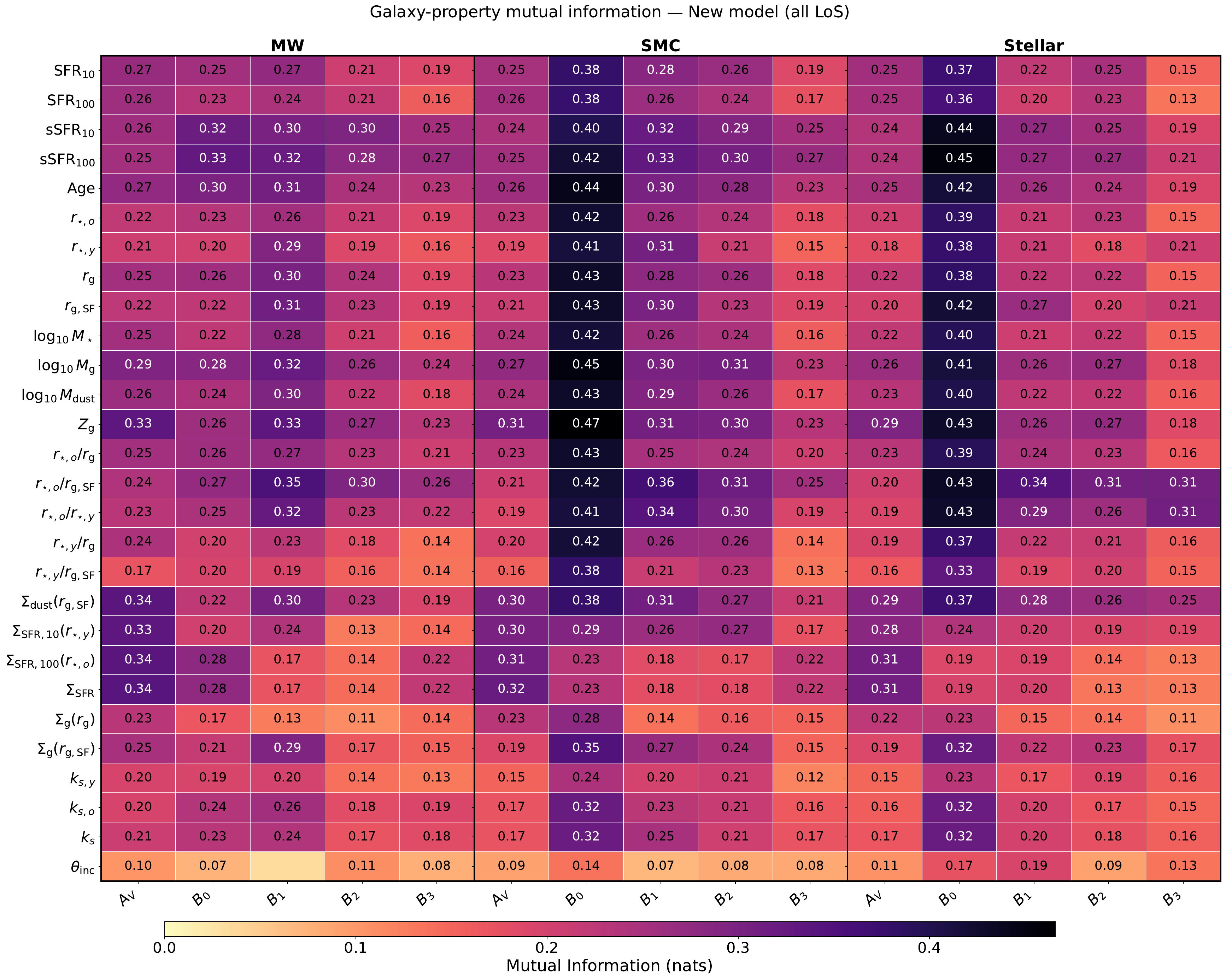}

\caption{
Same as Fig.~\ref{fig:correlations_MI} but for individual dust mixtures (MW, SMC, and stellar from left to right) separately.
}
\label{fig:correlations_MI_dust_mixes}
\end{figure*}

\cref{fig:correlations_dust_mixes,fig:correlations_MI_dust_mixes} demonstrate how the correlations with the parameters of this new model depend on dust mixture by considering the Spearman correlation coefficient and mutual information, respectively, for the MW, SMC and stellar dust mixtures. These are similar to \cref{fig:correlations_newmodel_dust_combined,fig:correlations_MI} from the main text, where we plot these correlations but averaged over dust mixture. 

\bibliographystyle{aasjournal}
\bibliography{bibliography}

\end{document}

%% file: definitions.tex


\def\be{\begin{equation}}
\def\ee{\end{equation}}
\newcommand{\code}[1]{\textsc{#1}}
\newcommand\quotesingle[1]{`{#1}'}
\newcommand\quotes[1]{``{#1}"}
\def\gsim{\lower.5ex\hbox{\gtsima}} 
\def\lsim{\lower.5ex\hbox{\ltsima}} 
\def\gtsima{$\; \buildrel > \over \sim \;$} 
\def\ltsima{$\; \buildrel < \over \sim \;$} \def\gsim{\lower.5ex\hbox{\gtsima}} 
\def\lsim{\lower.5ex\hbox{\ltsima}} 
\def\simgt{\lower.5ex\hbox{\gtsima}} 
\def\simlt{\lower.5ex\hbox{\ltsima}}

\def\msun{{\rm M}_{\odot}}
\def\lsun{{\rm L}_{\odot}}
\def\dsun{{\cal D}_{\odot}}
\def\fsun{\xi_{\odot}}
\def\zsun{{\rm Z}_{\odot}}
\def\msunyr{\msun {\rm yr}^{-1}}
\def\gdens{\msun\,{\rm kpc}^{-2}}
\def\sfrdens{\msun\,{\rm yr}^{-1}\,{\rm kpc}^{-2}}

\def\mum{\mu {\rm m}}
\newcommand{\angstrom}{\mbox{\normalfont\AA}}
\def\cc{\rm cm^{-3}}
\def\uflux{{\rm erg}\,{\rm s}^{-1} {\rm cm}^{-2} }

\def\fdust{\xi_{d}}
\def\fesc{f_{\rm esc}\,}
\def\td{\tau_{sd}}
\def\Sg{$\Sigma_{g}$}
\def\S*{$\Sigma_{\rm SFR}$}
\def\Ssfr{\Sigma_{\rm SFR}}
\def\Sgas{\Sigma_{\rm g}}
\def\Sstar{\Sigma_{\rm *}}
\def\Sesc{\Sigma_{\rm esc}}
\def\Srad{\Sigma_{\rm rad}}

\def\Dsolar{${\cal D}/\dsun$}
\def\Zsolar{$Z/\zsun$}
\def\DDsolar{\left( {{\cal D}\over \dsun} \right)}
\def\ZZsolar{\left( {Z \over \zsun} \right)}
\def\kms{{\rm km\,s}^{-1}\,}
\def\skms{$\sigma_{\rm kms}\,$}

\def\Scii{$\Sigma_{\rm [CII]}$}
\def\Sciimax{$\Sigma_{\rm [CII]}^{\rm max}$}
\def\CII{\hbox{[C~$\scriptstyle\rm II $]~}}
\def\CIII{\hbox{C~$\scriptstyle\rm III $]~}}
\def\OII{\hbox{[O~$\scriptstyle\rm II $]~}}
\def\OIII{\hbox{[O~$\scriptstyle\rm III $]~}}
\def\HH{\hbox{H$_2$}~} 
\def\HI{\hbox{H~$\scriptstyle\rm I\ $}} 
\def\HII{\hbox{H~$\scriptstyle\rm II\ $}} 
\def\CIion{\hbox{C~$\scriptstyle\rm I $~}}
\def\CIIion{\hbox{C~$\scriptstyle\rm II $~}}
\def\CIIIion{\hbox{C~$\scriptstyle\rm III $~}}
\def\CIVion{\hbox{C~$\scriptstyle\rm IV $~}}
\def\nhh{n_{\rm H2}}
\def\nhi{n_{\rm HI}}
\def\nhii{n_{\rm HII}}
\def\fhh{x_{\rm H2}}
\def\fhi{x_{\rm HI}}
\def\fhii{x_{\rm HII}}
\def\fd{f^*_{\rm diss}} 
\def\ks{\kappa_{\rm s}}

\def\cyan{\color{cyan}}
\definecolor{apcolor}{HTML}{b3003b}
\definecolor{afcolor}{HTML}{800080}
\definecolor{lvcolor}{HTML}{DF7401}
\definecolor{mdcolor}{HTML}{01abdf} 
\definecolor{cbcolor}{HTML}{ff0000}
\definecolor{sccolor}{HTML}{cc5500} 
\definecolor{sgcolor}{HTML}{00cc7a}